\documentclass[fleqn,usenatbib]{mnras}

\usepackage{newtxtext,newtxmath}

\usepackage[T1]{fontenc}
\usepackage{xcolor}
\usepackage[normalem]{ulem}

\DeclareRobustCommand{\VAN}[3]{#2}
\let\VANthebibliography\thebibliography
\def\thebibliography{\DeclareRobustCommand{\VAN}[3]{##3}\VANthebibliography}
\newcommand{\unitsSIv}{\ensuremath{\mathrm{km}\cdot\mathrm{s}^{-1}}}
\newcommand{\unitU}{\ensuremath{\mathrm{erg}\cdot \mathrm{cm}^{-3}}}
\newcommand{\unitLum}{\ensuremath{\mathrm{erg}\cdot\mathrm{s}^{-1}}}

\newcommand{\unitRBLR}{\ensuremath{\mathrm{R_{BLR}}}}

\usepackage{soul}
\usepackage{graphicx}	
\usepackage{amsmath}	
\graphicspath{{figures/} 
,{../../AllFigures/kappa/} 
,{../../AllFigures/analyses/}
}

\title[Cascade emissions from radio galaxies]{Compton-induced HE -- VHE $\gamma$-rays from pair cascade emissions in the SEDs of radio galaxies: NGC 1275}

\author[Ntshatsha, B{\"o}ttcher and Razzaque]{
M. Ntshatsha$^{1}$\thanks{E-mail: mfuphin95@gmail.com},
M. B{\"o}ttcher$^{2}$\thanks{E-mail: Markus.Bottcher@nwu.ac.za} and
S. Razzaque$^{1,3,4}$\thanks{E-mail: srazzaque@uj.ac.za}
\\
$^{1}$Centre for Astro-Particle Physics (CAPP) and Department of Physics, University of Johannesburg, Auckland Park 2006, South Africa\\
$^{2}$Centre for Space Research, North-Wet University, Potchefstroom, 2520, South Africa \\
$^{3}$ Department of Physics, The George Washington University, Washington, DC 20052, USA\\
$^{4}$ National Institute for Theoretical and Computational Sciences (NITheCS), Private Bag X1, Matieland, South Africa
}

\date{Accepted XXX. Received YYY; in original form ZZZ}

\pubyear{\the\year{}}

\begin{document}
\label{firstpage}
\pagerange{\pageref{firstpage}--\pageref{lastpage}}
\maketitle

\begin{abstract}
In the scheme of active galactic nuclei (AGNi), the blazar subclass is particularly bright in $\gamma$-rays. This is because of their emissions being beamed into our line of sight, as opposed to their misaligned parent population, radio galaxies. This work presents results of a Monte-Carlo code that propagates jet-collimated $\gamma$-rays through the magnetized AGN environment, leading to secondary cascade emissions from relativistic electron-positron pairs. The resulting multi-wavelength emissions are used to fit the broadband spectral energy distribution (SED) of the radio galaxy NGC 1275. Using known physical parameters of NGC 1275 from the literature, we find that cascade emissions from stochastic processes have the potential to reproduce the broadband SEDs of radio galaxies. We also find that the interplay between accretion disk and broad-line region (BLR) photons may provide insights into the production region(s) of $\gamma$-rays in AGNi. While dense accretion disk and BLR radiation fields are vital to the development of cascades, excessive densities lead to the suppression of cascade synchrotron emission and attenuation of \,TeV $\gamma$-rays. Furthermore, the magnetic field is important in shaping the cascade $\gamma$-ray spectrum.
\end{abstract}


\section{Introduction}
\label{sec:intro}
Active galactic nuclei (AGNi) host supermassive black holes (SMBHs) powering the radiative processes responsible for their electromagnetic (EM) emissions at all wavelengths \citep{urry1995unified, robson1996active, rivera2016agnfitter}. The spin of the black hole may launch collimated bipolar relativistic jets \citep{urry1995unified, dermer2016active, blandford2019relativistic}, where charged particles, confined to the jet by magnetic fields \citep{marti2019numerical}, can be efficiently accelerated. Gas and other material in-falling from the dusty torus form an accretion disk around the central black hole. Turbulent mutual interactions of this material lead to thermal emissions in the optical to ultraviolet (UV) spectrum \citep{shakura1973black, urry1995unified}. The combination of these processes contributes to the EM continuum emission from radio through to $\gamma$-rays. The gas and dust from the dusty torus can obscure ionization emission lines depending on an observer's line of sight to the central source. These emission lines appear broad, indicative of a region close to the black hole, the broad-line region (BLR), containing clouds of gas orbiting the central engine at very high velocities \citep[$\sim 10^3 - 10^4$\,\unitsSIv,][]{netzer1993dust, punsly2018revealing}. They are obscured in type 2 AGNi and prominent in type 1 AGNi such as blazars. The unified scheme for AGN of \citet{urry1995unified} suggests that all AGNi are essentially the same type of object. Their classifications are due to their orientation with respect to the observer's line of sight, which influences their apparent spectral properties. In the case of blazars, the observer's view falls within the cone of beamed emission from the jet. Bulk flows of plasma stream up the jet at relativistic speeds towards the observer, Doppler enhancing their emissions to high energies and luminosities, and reducing the variability time scales. This makes them ubiquitous sources of extra-galactic $\gamma$-rays. The jet axes of radio galaxies, on the other hand, are oriented at rather large angles with respect to the observer's line of sight \citep{sitarek2010timedependent}, such that the beamed emission from the jets is effectively invisible to the observer because of the aberration effects of special relativity \citep{rybicki1996radiative}. 

The vast majority of extragalactic sources dominating the $\gamma$-ray sky are blazars \citep{pushkarev2009jet, ajello2020fourth}. Improvements in $\gamma$-ray telescope sensitivities led to an increasing number of radio galaxies detected in high-energy (HE, $> 100$\,MeV) and very high-energy (VHE, $> 100$\,GeV) $\gamma$-rays \citep{sitarek2010timedependent, dermer2016active, tanada2018origins}. This, while supporting unification schemes that AGNi are fundamentally the same type of object, challenges the view that HE -- VHE emissions result from Doppler-boosted superluminal flows as seen by an observer whose line of sight aligns with the AGN jet.

Low-frequency emissions from the accretion disk and the BLR can intercept and absorb the HE -- VHE $\gamma$-rays from the jet through the $\gamma - \gamma$ pair production mechanism. In this mechanism, relativistic electron-positron (e$^\pm$) pairs are produced which can inverse-Compton (IC) scatter these same disk and BLR photons to HE and VHE. This process initiates a chain reaction, resulting in a cascade of e$^\pm$ pairs and secondary $\gamma$-rays. This framework was considered and pioneered in earlier works \citep[][]{sitarek2010timedependent, roustazadeh2010very, roustazadeh2011very, roustazadeh2012synchrotron}. In this paper, we extend the works of Roustazadeh \& B\"ottcher by including the direct accretion disk emission in the model and apply this framework to the radio galaxy NGC 1275.

In Section~\ref{sec:code} we provide details of the code used in this work and highlight the physical processes considered to result in the observed radiative output. In Section~\ref{sec:sim_agn} we give details of the simulated AGN. Section~\ref{sec:resuls} shows the results, discussing the effect of various parameters on the observed spectral energy distributions (SEDs). Based on the SED behaviour to various parameters, we also discuss a set of plausible parameter combinations capable of reproducing an observed flare of NGC 1275 in the same section. We summarize and conclude in Section~\ref{sec:discussion} and give our prospects of a future study in Section~\ref{sec:future}. Note that throughout this work all calculations are performed in the AGN rest-frame.

\section{Simulated interactions}
\label{sec:code}
The original Monte-Carlo (MC) code used in this work is described in \citet{roustazadeh2010very, roustazadeh2011very, roustazadeh2012synchrotron, roustazadeh2015time}. In \cite{ntshatsha2024compton} we made modifications to this code, referred to as the \emph{cascade code} from hereon, and demonstrated the plausibility of its radiative output. Here we describe the cascade code in detail and use it to fit the observed SED of the radio galaxy NGC 1275. From the vast works modelling AGN jets it is clear that these objects are complex and involve many physical processes that occur (quasi-)simultaneously. The cascade code focuses on $\gamma - \gamma$ absorption, external IC (EC) and synchrotron processes as the main mechanisms responsible for the observed emissions. The goal of the cascade code is to propagate individual $\gamma$-ray photons, considering all these mechanisms and generate EM emission resulting from Compton-supported e$^\pm$ pair cascades. 

\subsection{Code-setup}
\label{sec:codesetup}
\begin{figure*}
    \centering
    \includegraphics[width=0.8\linewidth]{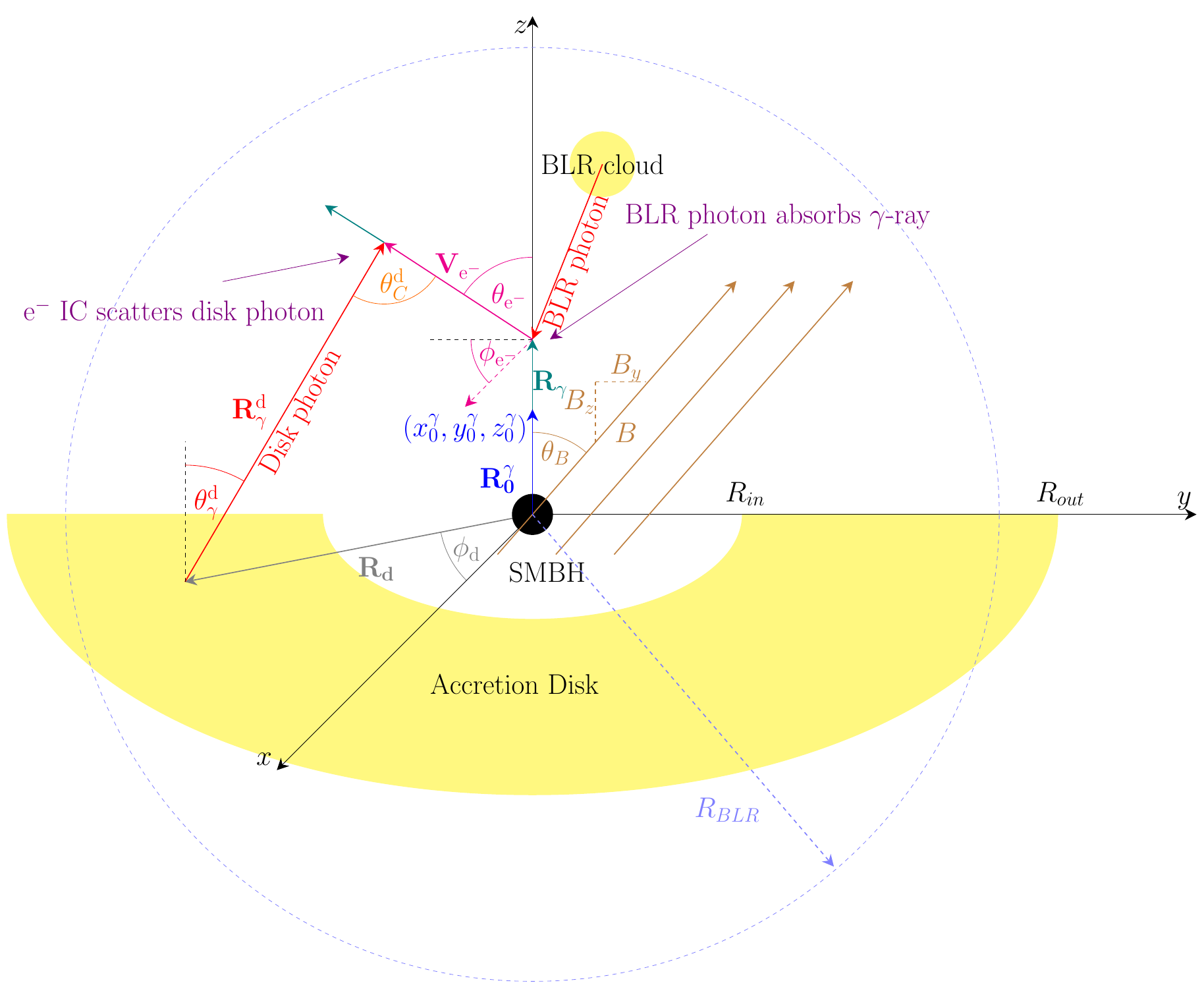}
    \caption{Geometry adapted from \citet{ntshatsha2024compton} describing the variables in the main text.}
    \label{fig:geometry}
\end{figure*}

The description of the cascade code and the physics considered in the simulation are given in this section. The geometry of the generic AGN environment we adopt is depicted in Figure 1 of \cite{roustazadeh2010very}. A jet axis is perpendicular to a geometrically thin \citet{shakura1973black} accretion disk (SS disk) with the radially dependent temperature profile \citep{shakura1973black, bottcher1997gammaray}
\begin{align}
\Theta(R) &\equiv k_B T(R)/(m_{\rm e}c^2) \nonumber\\&= 1.44\left(\frac{M}{\mathrm{M}_\odot}\right)^{-1/2}\left(\frac{\dot{M}}{\mathrm{M}_\odot \mathrm{yr}^{-1}}\right)^{1/4}
\left(\frac{R}{R_g}\right)^{-3/4}\left(1 - \sqrt{\frac{6R_g}{R}}\right)^{1/4}\,,
\label{eq:disk_temperature}
\end{align}
where $k_B, m_{\rm e}, T, M, \dot{M}, R, R_g \equiv GM/c^2$ are the Boltzmann's constant, electron rest-mass, temperature, the SMBH's mass, the mass accretion rate, the accretion disk radius, and the gravitational radius, with $G$ and $c$ being Newton's gravitational constant and the speed of light in vacuum, respectively. Energetic (0.5\,GeV -- 5\,TeV) intrinsic $\gamma$-ray photons are produced at a fixed point and propagate along the jet axis. We do not presume to know \emph{a priori} the mechanism(s) producing these primary $\gamma$-rays. They are assumed to result from non-thermal processes, hence following a power law ($\propto \epsilon^{-\alpha}_\gamma$) distribution in energy, consistent with \emph{Fermi} Large Area Telescope (LAT) observations of many blazars \citep[][]{abdo2009Fermi, abdo2010fermi}. The primary $\gamma$-rays are drawn from a power-law distribution with normalized photon energies, $\epsilon^*_\gamma \equiv E^*_\gamma/(m_{\rm e} c^2)$, between $10^{3}$ and $10^{7}$, where $E_\gamma^*$ is the primary $\gamma$-ray's energy. The $\gamma$-ray photon propagates and interacts with the low-frequency optical/UV photons from the BLR and the SS disk. For simplicity, we ignored infrared (IR) photons from the torus in this code. The BLR photon field is considered to be homogeneous and isotropic inside a volume of radius $R_{\rm BLR}$, and its energy density is given by the expression \citep[][]{roustazadeh2010very},
\begin{equation}
    u_{\rm BLR}(\epsilon, R) = u_0\delta(\epsilon - \epsilon_{\rm Ly\alpha})H(R_{\rm BLR}-R)\,,
    \label{eq:uBLR}
\end{equation}
where $\epsilon_{\rm Ly\alpha}$ is the dimensionless energy of the Lyman $\alpha$ ($\rm Ly\alpha$) emission line and $\delta(x), H(x)$ are the Dirac-delta and Heaviside functions, respectively. The accretion disk radiates thermal photons, with dimensionless temperature given by Equation~\eqref{eq:disk_temperature}, of the form:
\begin{equation}
    \frac{d^2 n_{\rm ph}}{dA_{\rm d} \, d\epsilon} = \frac{45}{2}\frac{\sigma_{SB}}{\pi^5}\frac{(m_{\rm e}c^2)^3}{ck_B^4}\frac{|\cos\theta^{\rm d}_\gamma|}{(R^{\rm d}_\gamma)^2} \frac{\epsilon^2}{\exp{[\epsilon/\Theta(R_{\rm d})]} -1}\,.
    \label{eq:disk}
\end{equation}
Here, $\sigma_{SB}$ is the Stefan-Boltzmann constant. In polar coordinates, if an accretion disk photon emanates from $(R_{\rm d}, \phi_{\rm d})$ on the disk (see Figure~\ref{fig:geometry}), where $R_{\rm d} = |{\bf R_{\rm d}}|, \phi_{\rm d}$ are the radius from the central engine in the plane of the disk and the azimuthal angle, respectively, then $R^{\rm d}_\gamma = |\bf R_{\gamma}^{\rm d}|$ is the distance from this point to the point of interaction with a $\gamma$-ray photon. The angle the disk photon makes with the normal to the disk is represented by $\theta^{\rm d}_\gamma$ and the dimensionless energy of low-frequency photons is denoted by $\epsilon$.

The cascade code tracks $\gamma$-rays individually, in 3-dimensional (3D) spatial coordinates. Each primary $\gamma$-ray is injected at some height $\mathbf{R}^\gamma_0 = (x^\gamma_0, y^\gamma_0, z^\gamma_0) = (0, 0, z^\gamma_0)$, where $z^\gamma_0$ is specified by the user (see Figure~\ref{fig:geometry} for a pictorial description of the geometrical variables). It travels along the jet-axis away from the central BH. Along their path, $\gamma$-rays may be absorbed by either a BLR or a disk photon, producing a relativistic e$^\pm$ pair. The total absorption coefficient, $\kappa_{\gamma\gamma}^{\rm tot}$, of the $\gamma$-ray is therefore calculated as the sum of contributions from the respective external radiation fields, namely $\kappa^{\rm d/BLR}_{\gamma\gamma}$, from the accretion disk/BLR, respectively, i.e., $\kappa_{\gamma\gamma}^{\rm tot} = \kappa_{\gamma\gamma}^{\rm d} + \kappa_{\gamma\gamma}^{\rm BLR}$. The absorption probability of a $\gamma$-ray is given by $1 - e^{(-\kappa_{\gamma\gamma}^{\rm tot} l_{\rm abs})}$, where $l_{\rm abs}$ is the distance the energetic photon will travel before it is absorbed, creating an e$^\pm$ pair. By equating this absorption probability to a uniform random number $\zeta_{\rm abs} \in (0,1)$ the cascade code determines $l_{\rm abs}$. If $\bf k_\gamma$ is the unit vector pointing in the direction of the $\gamma$-ray, then the vector ${\bf R}_\gamma = l_{\rm abs}{\bf k_\gamma}$ tracks the $\gamma$-ray. At $l_{\rm abs}$, the $\gamma$-ray can either be absorbed by a disk or BLR photon. The code determines this by simply drawing a uniform random number $\zeta_{\rm sel} \in (0,1)$: if $\zeta_{\rm sel} < \kappa_{\gamma\gamma}^{\rm d}/\kappa_{\gamma\gamma}^{\rm tot}$, the absorbing photon is a disk photon, otherwise it is a BLR photon. All of the energy and momentum of the $\gamma$-ray is imparted to the e$^\pm$ pair. The relativistic electron (e$^-$) and positron (e$^+$), initially travelling in the direction of the $\gamma$-ray, are deflected by the ambient magnetic field. The e$^\pm$ trajectories are determined by solving Euler-Lagrange equations where the magnetic field is uniform in the $y-z$ plane. The velocities from these equations form the components of the velocity vector, $\bf V_{\rm e^-}$, which tracks an electron ($\bf V_{\rm e^+}$ for the positron). In this process, the pairs lose energy by emitting synchrotron photons and by IC scattering BLR and disk photons to high frequencies. In case of an IC scattering, the position vector ${\bf R}_{0}^\gamma$ updates to the IC coordinates, resetting $\bf R_\gamma$ (follow green arrow vector of Figure~\ref{fig:geometry}) and the secondary $\gamma$-ray is tracked in the same way as the primary $\gamma$-ray. Of course, in principle, these also IC-scatter the synchrotron photons in the synchrotron self-Compton (SSC) scenario. The cascade synchrotron luminosity goes as $L_{\rm sy} \sim 4 \pi d_L^2 \nu F_{\nu}^{\rm sy}$, where $d_L = 78$\,Mpc \citep{reynolds2021probing} is the luminosity distance of NGC 1275 and $\nu F_\nu^{\rm sy}$ is the average cascade synchrotron flux emitted within the polar angular bin $14\,\degr \leq \theta \leq 17\,\degr$. At the highest simulated magnetic field of $B = 100$\,mG, BLR energy density of $u_{\rm BLR} = 50\times 10^{-3}$\,\unitU\ and injection height of $z_0^\gamma = 0.8$\,\unitRBLR, the maximum simulated cascade synchrotron energy density is then $u_{\rm sy} = L_{\rm sy} / (4\pi c R_{\rm BLR}^2) \approx 54\times 10^{-3}$\,\unitU. In all our simulations, we considered BLR energy densities in the range of $\sim (5 - 50)\times 10^{-3}$\,\unitU\ and with the fixed accretion disk luminosity of $\sim 2\times 10^{43}$\,\unitLum, the energy density of the accretion disk radiation field is fixed at $50\times 10^{-3}$\,\unitU. These estimates show that the cascade synchrotron energy density becomes comparable to the external radiation fields at high magnetic fields, thus making SSC important. However, at high energy densities of the external radiation fields, they can conspire to suppress the $u_{\rm sy}$ (see the discussion in Section~\ref{sec:BLRdensity}). As a pilot study, we have limited our cascade code to external Comptonization of only the BLR and SS disk radiation fields. A description of the e$^\pm$ tracking routine is given in Section \ref{sec:paircascade}. The cascade code tracks these secondary $\gamma$-rays recursively using the routine just described above. This results in a cascade of e$^\pm$ and second-generation $\gamma$-ray photons. A secondary photon is tracked until it escapes the AGN volume ($r_{\rm abs} > R_{\rm BLR}$, where $r_{\rm abs}$ is the distance from the SMBH to $\gamma$-ray absorption (or, equivalently, e$^\pm$ pair creation point)) or its energy falls below the \emph{Fermi} range (we set the cutoff to $\epsilon_{\gamma,\rm cut} = 10$ or $\sim 5$\,MeV). An electron/positron is tracked until its Lorentz factor falls below the cut-off value of $\gamma_{\rm cut} = 100$, or it escapes the AGN volume.

To simulate $\gamma$-ray propagation in an AGN the code requires the following input parameters: The magnetic field $\mathbf{B} = (B_x, B_y, B_z) = (0, B_y, B_z)$ (this field is assumed uniform and static), the size of the broad-line region $R_{\rm BLR}$, the primary $\gamma$-ray spectral index $\alpha$, BLR energy density $u_{\rm BLR}$, accretion disk luminosity $L_{\rm d}$, primary $\gamma$-ray injection height $z^\gamma_0$, supermassive black-hole mass $m_{\rm BH}$ and the angle $\theta_B$ that the magnetic field makes with the jet axis. Not all these parameters are known empirically. In the case of NGC 1275 we set the primary $\gamma$-ray spectral index $\alpha = 2.5$ \citep[consistent with][]{roustazadeh2010very, godambe2024very}, the BLR radius and accretion disk luminosity to $R_{\rm BLR}=10^{16}$\,cm and $L_{\rm d} = 1.88\times 10^{43}$\,\unitLum, respectively \citep[][ consistent with the $R_{\rm BLR}\propto \sqrt{L_{\rm d}}$ scaling relation]{roustazadeh2010very, ghisellini2009canonical}. The mass of the SMBH has been set to $m_{\rm BH} = 10^{8} \, {\rm M}_{\odot}$ and the angle of the magnetic field to $\theta_B \equiv \arctan{(B_y/B_z)} = 11\,\degr$ with respect to (w.r.t) the jet-axis, following the previous works of \citet{roustazadeh2010very}. The remaining parameters, namely $B, u_{\rm BLR}$ and $z^\gamma_0$ are left as free parameters. In a post-processing analysis we generate SEDs for various viewing angles. These are compared with the observed SED of NGC 1275, with our selection of viewing angular bins guided by characterizing beaming angles estimated in the literature.

\subsection{Target photon fields}
\label{sec:softph}
Morphologically, AGNi are complex, anisotropic structures. In the AGN rest-frame, where we do our calculations, the broad emission line radiation field is essentially isotropic inside the BLR. The $\gamma$-ray photons in the jet can be intercepted by BLR photons or direction-dependent accretion disk photons. The absorption coefficient of the intercepting photon is strongly influenced by the height of the $\gamma$-ray above the SMBH. While the BLR radiation field is assumed to be homogeneous, it is more important at greater heights with the accretion disk's dominance being more important at lower heights. However, in reality the $\gamma-\gamma$ interaction is ultimately stochastic \citep[see discussion in][]{ntshatsha2024compton}. The $\gamma-\gamma$ absorption is calculated using the full cross-section formula
\begin{equation}
    \sigma_{\gamma\gamma}(\beta) = \frac{3}{16} \, \sigma_T \, (1-\beta^2) \, \left(\left[3-\beta^4\right]\ln{\frac{1+\beta}{1-\beta}} -2\beta\left[2-\beta^2\right]\right)\,,    
\end{equation}
where $\sigma_T$ is the Thomson cross-section, $\beta \equiv \sqrt{1-2/(\epsilon\epsilon_\gamma(1-\mu))}$ and $\epsilon, \epsilon_\gamma, \mu$ are dimensionless target photon energy, $\gamma$-ray energy and the polar angle cosine of interaction between the target photon and the $\gamma$-ray \citep{jauch1976theory}, respectively.

\begin{figure}
	\includegraphics[width=0.95\columnwidth]{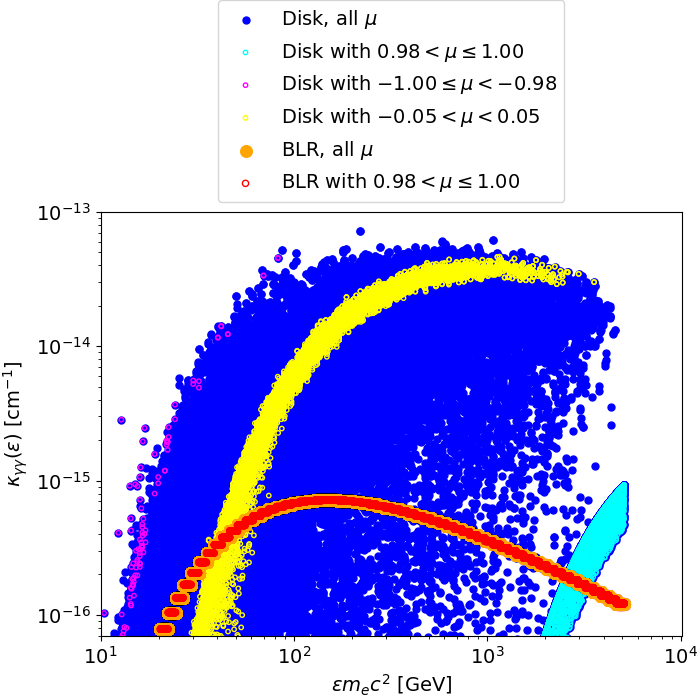}
    \caption{Scatter plot of absorption coefficient values to individual $\gamma$-rays produced in the MC simulation. 
    \emph{Orange} dots correspond to BLR photons absorbing $\gamma$-rays travelling in all polar directions, with \emph{red} circles representing those travelling in the forward direction. The \emph{blue} dots are accretion disk photons absorbing $\gamma$-rays  travelling in all polar directions. The \emph{cyan, yellow} and \emph{magenta} circles are absorption events by disk photons of $\gamma$-rays travelling in forward, edge-on and backward directions, respectively. In both disk and BLR cases, the absorption events occur for $\gamma$-rays travelling at all azimuthal directions.
    }
    \label{fig:opacity}
\end{figure}

Accretion disk absorption coefficient is given by
\begin{align}
\kappa^{\rm d}_{\gamma\gamma}(\epsilon_\gamma, \mu^{\rm d}_{\gamma\gamma}) &= \int_{R_{in}}^{R_{out}} R_{\rm d} \,dR_{\rm d}
 \int_0^{2\pi} \,d\phi_{\rm d} \nonumber\\
 &\times\int_0^\infty \,d\epsilon
         \frac{d^2n_{\rm ph}}{dA_{\rm d}d\epsilon}
         (1-\mu^{\rm d}_{\gamma\gamma})
         \sigma_{\gamma\gamma}(\epsilon_\gamma, \epsilon, \mu^{\rm d}_{\gamma\gamma})
\label{eq:kappaDisk}
\end{align}
where $R_{in} = 6R_g$, $R_{out} = 10^3R_{in}$ are the inner and outer disk radii, respectively, $\phi_{\rm d}$ is the azimuthal angle in the disk plane and $\mu^{\rm d}_{\gamma\gamma}$ is the interaction polar angle cosine between a disk photon and a $\gamma$-ray. Figure~\ref{fig:opacity} shows a scatter plot of $\gamma$-ray attenuation events taking place in the BLR and SS disk radiation fields, in a simulation where $B = 50$\,mG, $u_{\rm BLR} = 50\times 10^{-3}$\,\unitU\ and $z^\gamma_0 = 0.8$\,\unitRBLR. In the simulations, whenever a $\gamma$-ray's absorption coefficient is calculated, its direction and energy are noted. Figure~\ref{fig:opacity} shows these attenuation events for all azimuthal angles. Of the $\gamma$-rays absorbed by disk photons, those travelling in all polar angles are shown in \emph{blue} dots. We then isolate polar cosine angular bins ($\mu = \cos\theta$) corresponding to $0\,\degr \leq \theta \leq 10\,\degr$ ($0.98< \mu \leq 1$, \emph{cyan circles}), $87\,\degr \leq \theta \leq 93\,\degr$ ($-0.05 < \mu < 0.05$, \emph{yellow circles}) and  $170\,\degr \leq \theta \leq 180\,\degr$ ($-1 \leq \mu < -0.98$, \emph{magenta circles}) representing the forward, edge-on and backward directions, respectively. Absorption by BLR photons is represented by \emph{orange} dots, with absorption taking place in the $0.98 < \mu \leq 1$ bin indicated by \emph{red} circles.
Note that the range of $\kappa_{\gamma\gamma}$ values due to disk photons is rather large compared to that due to BLR photons. This is due to the assumption that the BLR radiation field is homogeneous throughout the AGN. This ensures that the $\kappa_{\gamma\gamma}^{\rm BLR}$ is the same for all photons of the same energy, independent of direction and the location where the $\gamma - \gamma$ absorption takes place. Though the cascade code can be generalised to include additional emission lines from the BLR, for our simulations we have assumed it to be dominated by the $\rm Ly\alpha$ emission line, which is a sufficient representation of the BLR \citep[see][]{stern2014mystery}. The figure also shows that $\gamma$-rays with energies $\gtrapprox 20$ \,GeV up to \,TeV energies are highly susceptible to absorption by optical/UV photons. By contrast, the accretion disk radiation field is anisotropic; thus some disk photons interact with $\gamma$-rays at angles more favourable than others. In Figure~\ref{fig:opacity}, the $\mu$-dependent absorption patterns by disk photons highlight the efficiencies of the interaction angles between $\gamma$-rays and disk photons. In forward directions ($0.98< \mu \leq 1$), the interaction angles are unfavourable since these $\gamma$-rays and disk photons happen to be travelling in similar directions; thus, to meet the pair-production threshold, the $\gamma$-ray energies have to be very high. Lower-energy $\gamma$-rays that meet the pair-production threshold happen to be travelling in the backward direction and meeting disk photons in head-on collisions. Also, the $\cos{\theta^{\rm d}_\gamma}$ term appearing in $\,d^2n_{\rm ph}/\,dA_{\rm d}\,d\epsilon$ in Equations~\eqref{eq:disk} and \eqref{eq:kappaDisk}, gives $\kappa_{\gamma\gamma}^{\rm d}$ a dependence on the height above the disk. This results in a broad range of $\kappa_{\gamma\gamma}^{\rm d}$ values, even for $\gamma$-rays of identical energies, as seen in Figure~\ref{fig:opacity}. The low-energy boundary in Figure~\ref{fig:opacity} is due to the pair-production threshold $\epsilon_\gamma >  2/[\epsilon(1-\mu^{\rm d}_{\gamma\gamma})]$.

\subsection{Secondary $\gamma$-rays from e$^\pm$ pairs}
\label{sec:paircascade}
Attenuation described in Section~\ref{sec:softph} above occurs due to the $\gamma-\gamma$ pair creation mechanism. The prescription governing their creation is derived and demonstrated in the work of \citet{bottcher1997pair}. Supposing the absorbed $\gamma$-ray's normalized energy is $\epsilon_{\gamma}$, neglecting the much smaller energy of the target photon, the cascade code then samples the electron and positron's Lorentz factors $\gamma_{\rm e^-}$ and $\gamma_{\rm e^+} = (\epsilon_{\gamma} - \gamma_{\rm e^-})$, respectively, from the pair spectrum described in that work. The designation of their values is determined by a random number $\zeta_\pm \in (0,1)$. If $\zeta_\pm < 0.5$, an electron's Lorentz factor, $\gamma_{\rm e^-}$, is drawn from the pair spectrum and the positron's will be $\gamma_{\rm e^+} = \epsilon_{\gamma} - \gamma_{\rm e^-}$. Otherwise (if $\zeta_\pm \geq 0.5$), a positron's $\gamma_{\rm e^+}$ is drawn and the electron's is $\gamma_{\rm e^-} = \epsilon_{\gamma} - \gamma_{\rm e^+}$. The inverse-Compton mean-free-path length in the accretion disk radiation field is given by
\begin{align}
 \lambda_{C,\rm d}^{-1} &= \int_0^{\infty} 
                         \,d\epsilon
                    \int_{-1}^1
                         \,d\mu_C \nonumber\\
                    &\times \int_0^{2\pi}
                         \,d\phi_{\rm d}
                         n_{\rm ph}(\epsilon, \Omega)
                         (1 - \beta\mu_C)
                         \sigma_C(\epsilon, \gamma, \mu_C)
\label{eq:MFP_compton} 
\end{align}
where $n_{\rm ph}$ is the number density of the accretion disk radiation field given in Equation~\eqref{eq:disk} and $\Omega$ is the solid angle. The Compton cross-section is given by
\begin{equation}
 \sigma_C = \frac{2\sigma_{\rm e}} {x}
                       \left[
                            \left(
                                1 - \frac{4}{x} - \frac{8}{x^2}
                            \right)\ln{(1 + x)}
                                + \frac{1}{2} + \frac{8}{x}
                                - \frac{1}{2(1 + x)^2}
                       \right]\,,
\label{eq:Compton_crosssection}
\end{equation}
where $x = 2 \epsilon\gamma_{\rm e^\pm} (1 - \beta_C\mu_C)$, $r_{\rm e} \equiv e^2/(m_{\rm e}c^2)$ is the classical electron radius, $e$ is the elemental charge, $\sigma_{\rm e} = \pi r_{\rm e}^2$, $\gamma_{\rm e^\pm}$ is the incoming particle's Lorentz factor, $\beta_C\equiv \sqrt{1-\gamma_{\rm e^\pm}^{-2}}$, $\epsilon$ is the dimensionless disk photon energy and $\mu_C = \cos\theta_C^{\rm d}$ is the interaction polar angle cosine between the particle and disk photon (see Figure~\ref{fig:geometry}).

\begin{figure}
	\includegraphics[width=0.95\columnwidth]{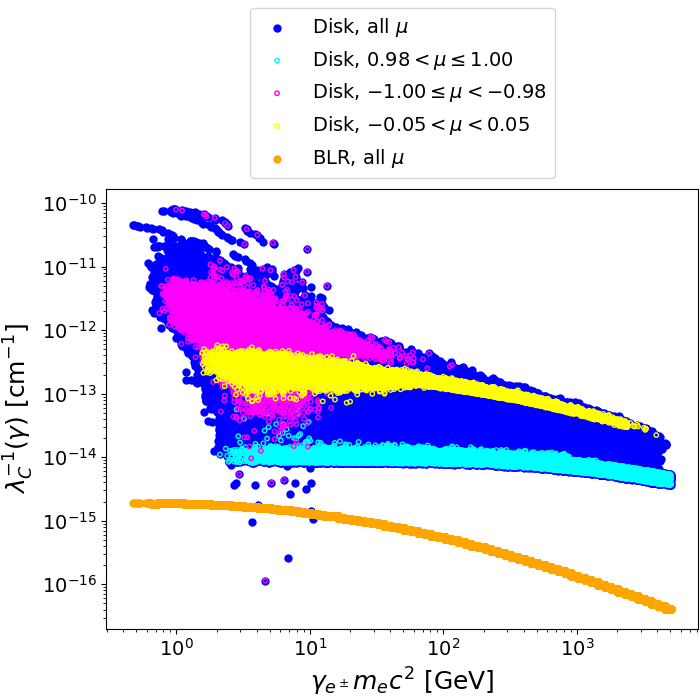}
	\caption{Scatter plot of inverse Compton mean-free path lengths ($\lambda^{-1}_C$) of individual e$^\pm$ particles produced in the MC simulation. 
    \emph{Orange} and \emph{blue} dots are the $\lambda^{-1}_C$ values of e$^\pm$'s upscattering BLR and accretion disk photons, respectively, in all polar directions. \emph{Cyan, yellow} and \emph{magenta} circles disk photons represent IC scattering in forward, edge-on and backward directions, respectively. These IC scattering events occur for e$^\pm$ travelling in all azimuthal directions, for both disk and BLR cases.
    }
	\label{fig:invMFP}
\end{figure}

The created particles, initially travelling in the direction of the absorbed $\gamma$-ray, are deflected by the magnetic field in the AGN environment. Their deflection then reduces their energy due to synchrotron radiation. In addition to synchrotron losses, they also lose energy to IC collisions with BLR and SS disk photons. Figure~\ref{fig:invMFP} is a scatter plot of the IC mean-free path length, $\lambda^{-1}_C$, in a simulation where $B = 50$\,mG, $u_{\rm BLR} = 50\times 10^{-3}$\,\unitU\ and $z^\gamma_0 = 0.8$\,\unitRBLR. In the figure, $\gamma_{\rm e^\pm}$ collectively denotes Lorentz factors of electrons and positrons. The sum $\lambda^{-1}_{C,\rm tot} = \lambda^{-1}_{C, \rm BLR} + \lambda^{-1}_{C, \rm d}$, of the inverse Compton mean-free-path lengths in the BLR and accretion disk, respectively, is useful in determining the distance to IC scattering. As an analogue to $l_{\rm abs}$, the distance $l_{\rm IC}$ to an IC upscattering is calculated from $l_{\rm IC} = \lambda_{C, \rm tot}\ln{(1-\zeta_{\rm sc})}$, where $\zeta_{\rm sc} \in (0,1)$ is a random number. When the electron or positron IC scatters an external photon, the upscattered photon is selected by comparing the random number $\zeta_{C,sel} \in (0,1)$ to $\lambda^{-1}_{C, \rm d}/\lambda^{-1}_{C, \rm tot}$. If $\zeta_{C, sel}$ is less than this ratio, then a disk photon is upscattered, otherwise it is a BLR photon. IC scattering events occur at all azimuthal directions. Events occurring in all polar directions are shown by \emph{blue} dots in the case of IC scattered disk photons and \emph{orange} dots in the case of IC scattered BLR photons. A trend is not readily obvious from the $\lambda^{-1}_{C,\rm d}$ for events occurring at all $\mu$, except for a steady overall decline towards higher energies due to the Klein-Nishina (KN) effect. The individual $\lambda^{-1}_{C,\rm d}$ values populate a wide range of values even for particles of similar energy. By contrast, the curve traced out by the simple BLR model has a one-to-one correspondence and suggests that $\lambda^{-1}_C$ is quasi-constant for particles with $\gamma_{\rm e^\pm} m_{\rm e}c^2 < 10$\,GeV, decreasing (due to the KN effect) by about 1.5 orders of magnitude up to \,TeV energies. This bijective behaviour is a consequence of our assumption that the BLR photon field is homogeneous and mono-energetic. To discern a basic trend from IC scattered disk photons, $\mu$ is sliced into the same polar angular bins described in Subsection~\ref{sec:softph}. The e$^\pm$ pairs IC scattering disk photons in the forward direction have a quasi-constant $\lambda_C^{-1}$ over a wide range of energies ($\sim 2$\,GeV to few \,TeV). At edge-on angles, the $\lambda_{C,\rm d}^{-1}$ starts to turn over, remaining quasi-constant below $\sim 100$\,GeV and slowly decreasing above this energy up to TeV energies. Pairs travelling in the backward direction tend to populate the \,GeV to sub 100\,GeV region. The density of the $\lambda^{-1}_{C, \rm d}$ scatter plot of e$^\pm$ pairs with energy below $\sim 10$\,GeV and quasi-constant $\lambda^{-1}_{C,\rm BLR}$ below this energy suggests that these particles are efficient at upscattering optical/UV photons.

\begin{figure}
    \centering
    \includegraphics[width=0.95\linewidth]{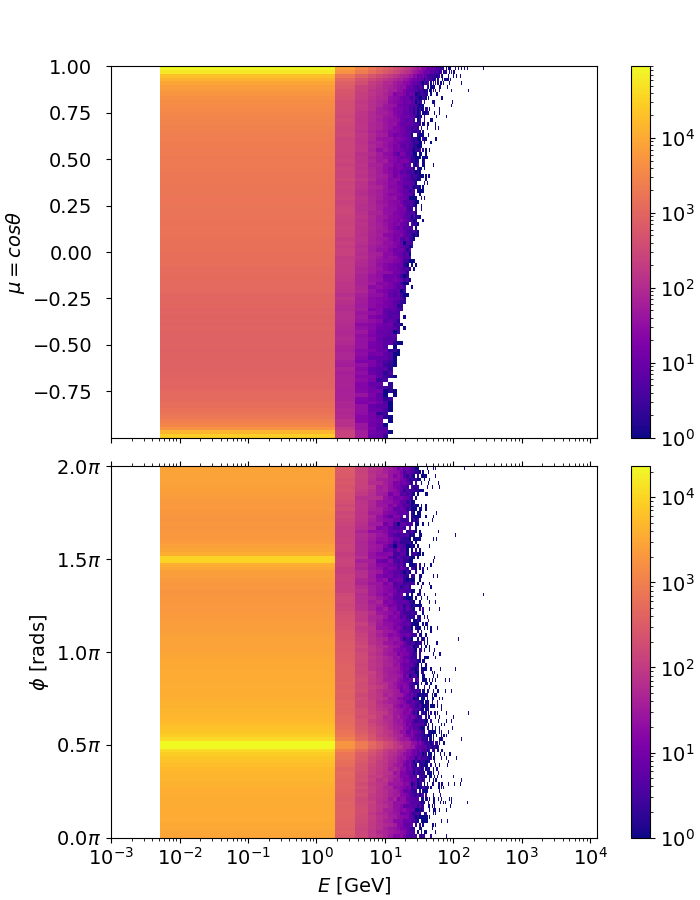}
    \caption{\emph{Upper panel}: 2D Histogram plot of angle cosine polar angles versus cascade $\gamma$-ray energy. \emph{Lower panel}: 2D Histogram plot of azimuthal angle versus cascade $\gamma$-ray energy.
    }
    \label{fig:angular_distr}
\end{figure}

Figure~\ref{fig:angular_distr} is a 2 dimensional (2D) histogram plot with a heatmap indicating cascade $\gamma$-ray counts.  The figure shows the angular distribution of the cascade emission generated with a parameter combination of $(B, u_{\rm BLR}, z^\gamma_0, \theta_B) = (50 \,{\rm mG}, 50\times 10^{-3}\,\unitU, 0.3\,\unitRBLR, 11\degr)$. We note that while the e$^\pm$ pairs are isotropized, their emission is partially isotropic in the sense that at all $\gamma$-ray energies, they escape mostly uniformly in the azimuthal direction, with a slight bias in the direction perpendicular to the plane of the magnetic field, as shown by the lower panel. Deep within the AGN, mostly all VHE $\gamma$-rays are absorbed. In general, cascade $\gamma$-rays with energies up to $\lesssim 10$\,GeV are essentially uniformly distributed in the polar direction (upper panel). Fewer photons escaped with energies above $\sim (1 - 10)$\,GeV (both panels), with the more energetic ones favouring the forward direction $\mu \sim 1$ (consistent with Figure~\ref{fig:invMFP}, \emph{also see Figure~\ref{fig:angular_distrz0.8}, where $z_0^\gamma = 0.8\,\unitRBLR$, for a clearer illustration of this point}). These are HE--VHE $\gamma$-rays that have experienced relatively low $\gamma-\gamma$ absorption rates. They trace out the curved feature in the upper panel above $\sim 10$\,GeV. Furthermore, although the analysis presented in Figure~\ref{fig:angular_distr} was generated with a specific parameter combination, the same general trend is observed for any $\theta_B$ and $(B, u_{\rm BLR}, z^\gamma_0)$-combination. We further note that, given the slight bias of the cascade to directions perpendicular to ${\bf B}$, we have kept an equal statistical weighting to all the azimuthal directions, in our analysis. This may be justified in cases of toroidal-oriented $B$ fields, which are more realistic representations of AGNi.
\section{Simulated AGN: NGC 1275}
\label{sec:sim_agn}
In this study, we simulated $\gamma$-rays propagating in the AGN environment of NGC~1275. With different parameter combinations, we generated cascade broadband SEDs from the simulation outputs. This simulated source has a Fanaroff-Riley I (FR I) jet structure. It is at redshift $z = 0.0176$ located in the centre of the Perseus cluster. Earlier studies of its jet morphology constrained its jet-axis angle w.r.t. our line of sight to be $\theta \sim 30\,\degr - 55\,\degr$ \citep{vermeulen1994discovery, walker1994detection, asada2006expanding}, while \citet{krichbaum1992evolution} estimated a minimum angle of $\theta \sim 14\,\degr$. Recent works by \citet{godambe2024very} used reports of the source's latest flaring events to constrain this angle to a maximum of $\theta \sim 17$\,\degr. During these flares, \,TeV $\gamma$-rays from this source were detected in the period between December 2022 and January 2023 \citep{godambe2024very, cao2024detection}. \citet{godambe2024very} used quasi-simultaneous data from the \emph{Swift} Ultraviolet and Optical Telescope (UVOT), \emph{Swift} X-ray Telescope (XRT), \emph{Fermi}-LAT and the Major Atmospheric Cherenkov Experiment (MACE) to construct broadband SEDs with flux spanning from optical/UV through to VHE $\gamma$-rays. They divided the flare period into three epochs, which they denote as P1, P2 and P3. These labels correspond to flares during the nights of 21 December 2022 and 10 January 2023, and the intermediate state between these flares, respectively. In our SED plots, we designate the \emph{blue}, \emph{orange}, \emph{green} and \emph{red} dots to correspond to \emph{Swift}-UVOT, \emph{Swift}-XRT, \emph{Fermi}-LAT and MACE data points, respectively. X-ray data (\emph{orange} points) corresponds to fluxes observed on Modified Julian Dates (MJDs) 59933.9 (\emph{upward pointing triangles}) and 59935.9 (\emph{downward pointing triangles}) and the average flux of these days (\emph{round} dots), as described in \citet{godambe2024very}. During the observations, the source was not spatially resolved. It is, therefore, worth noting that the optical/UV and X-ray fluxes are subject to contamination from the host galaxy, accretion-disk -- corona system, and possibly also the intracluster medium \citep[see][]{godambe2024very}. Hence, also given the misaligned nature of the source it is not entirely clear that the photon flux in these two frequency bands is dominated by emission from the jet. Thus, the data in these frequency bands are effectively upper limits for the cascade emission.
\section{Results}
\label{sec:resuls}
The cascade results presented in this work are compared to the P1 data of \citet{godambe2024very}, that is the night of 21 December 2022. In addition to their constraints on $\theta$, they interpreted NGC 1275's broadband SEDs using a one-zone SSC model and constrained its Doppler factor to the range (2 -- 3.5). This is a fundamental difference from this work, as all calculations performed by the cascade code are done in the AGN rest-frame, thus eliminating the need for Doppler boosting altogether. All cascade SEDs are of photons escaped from the polar angle cosine bin $0.95 < \mu\equiv\cos\theta < 0.97$ corresponding to viewing angles in the range $14\,\degr < \theta < 17\,\degr$, in line with $\theta$ constrains in the literature. Note that since NGC 1275's jet is misaligned, we have no face-on view of its jet. We have thus followed \citet{roustazadeh2012synchrotron} to calibrate the forward cascade emission by normalising it to the jet luminosity of the typical blazar 3C 279. This section explores an SED parameter study by changing the magnetic field, BLR energy density and primary photon injection height.
\subsection{Magnetic field}
\label{sec:Bfield}
\begin{figure}
	\centering
	\includegraphics[width=0.95\linewidth]{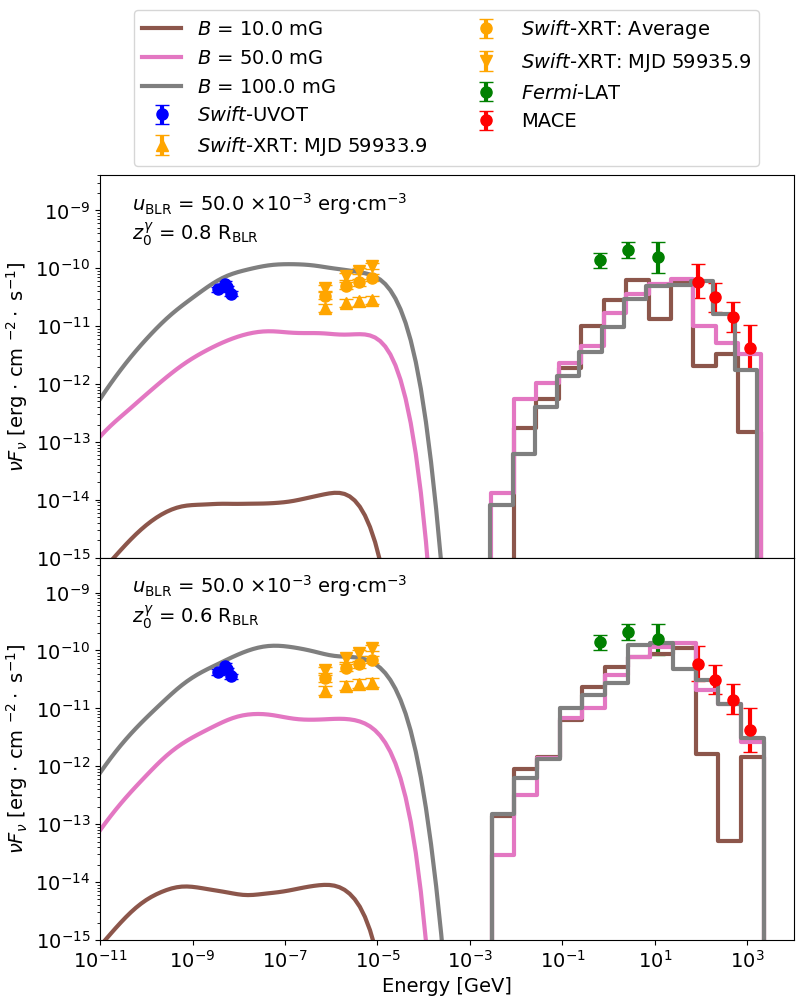}
	\caption{Cascade SEDs of simulations run at different magnetic field values. In these simulations, the BLR energy density was fixed to $u_{\rm BLR} = 50\times 10^{-3}$\,\unitU\ and primary photons were injected at a height of 0.8\,\unitRBLR\ and 0.6\,\unitRBLR\ in the \emph{upper} and \emph{lower} panels, respectively.
    Note that in this and subsequent figures, the remaining AGN parameters stay fixed as mentioned in Section~\ref{sec:codesetup}, namely $\alpha = 2.5,\ L_{\rm d} = 1.88\times 10^{43}\,\unitLum,\ R_{\rm BLR} = 10^{16}\,{\rm cm},\ m_{\rm BH} = 10^{8}\,{\rm M_\odot},{\rm\ and\ }\theta_B = 11\,\degr$. The SED data points in this and subsequent figures are described in Section~\ref{sec:sim_agn} and in the legend of this figure. 
    }
	\label{fig:SED_Bu50z0.8_varB_mu0.94-0.98}
\end{figure}

To study the resulting cascade emission with changing magnetic field strength, we fix the BLR energy density and primary $\gamma$-ray injection height. By changing the magnetic field the cascade SED behaviour was similar across different ($u_{\rm BLR}, z^\gamma_0$)-combinations (see Figure~\ref{fig:SED_Bu50z0.8_varB_mu0.94-0.98}, where the $z^\gamma_0$ is 0.8\,\unitRBLR\ and 0.6\,\unitRBLR\ in the upper and lower panels, respectively). The cascade synchrotron peak flux increases with increasing magnetic field $\left(\nu F_{\nu}^{\rm sy, peak} \propto B^2\right)$ which is to be expected. The IC flux also rises with increasing $B$. However, it abruptly saturates to a maximum. We know from \citet{roustazadeh2010very} that the cascade is especially sensitive to transverse magnetic fields. In our simulation runs, the magnetic field is oriented at an angle of $\theta_B = 11\,\degr$ w.r.t the jet-axis, resulting in a rather modest transverse $B$ component. At this $\theta_B$, the IC component of the cascade flux reaches saturation at $B$ strengths in the order of multiple tens of \,mG. This rapid saturation of IC flux is interpreted as the isotropization of the e$^\pm$ pairs in the AGN environment by this transverse $B$-field \citep{roustazadeh2010very}, as shown in Figure~\ref{fig:angular_distr}.

In addition to the isotropization of pairs, in the case of a uniform $B$-field, this work also notes that the magnetic field has an effect on the shape of the $\gamma$-ray spectrum. Low $B$ strengths result in a narrow band peaking at the HE-end of the spectrum, and broadening towards the VHE-end as $B$ increases. Our diagnosis of this is that the gyro-radius $(r_L\propto \gamma_{\rm e^\pm} B^{-1})$ of e$^\pm$ pairs with energy $E_{\rm e^\pm} = \gamma_{\rm e^\pm} m_{\rm e}c^2$ is large at low magnetic fields, leading to inefficient deflection of the pairs. Additionally, comparing the relatively short IC cooling length \citep{rybicki1996radiative, dermer2009high, roustazadeh2010very},
\begin{equation}
    D_{\rm IC} =\frac{
        3m^2_{\rm e}c^4
    }{
        4\sigma_{\rm T}
    } \frac{
        4\pi cR^2_{\rm ext}
    }{
        \tau_{\rm ext} L_{\rm d}
    } \frac{1}{\beta^2 E_{\rm e^\pm}} 
    \approx 1.6\times 10^{12} E^{-1}_{\rm TeV} \tau^{-1} L_{43}^{-1}R_{16}^2\,\quad\,{\rm cm},
\label{eq:Dic}
\end{equation}
to the Larmor radius, 
\begin{equation}
    r_{\rm L} = \frac{E_{\rm e^\pm}}{eB} \approx 3.3\times 10^{12} E_{\rm TeV}B_{-3}^{-1}\quad\,{\rm cm},
\label{eq:rL}
\end{equation}
gives the typical deflection angle of the pairs to be
\begin{equation}
    \theta_{\rm IC} = D_{\rm IC}/r_{\rm L} \sim 29\,\degr E_{\rm Tev}^{-2}L^{-1}_{43}\tau^{-1}R^2_{16}B_{-3},
\label{eq:cooling_lengthIC}
\end{equation}
between IC collisions, where e$^\pm$ pairs with the highest energies impart a large fraction of their energy to ambient photons in the forward direction. In these estimates (Equations~\eqref{eq:Dic},~\eqref{eq:rL},~\eqref{eq:cooling_lengthIC}), a relativistic electron, in the Thomson regime, cools by IC scatterings in an external radiation field with parameterised luminosity $\tau_{\rm ext}L_{\rm d}$, where $L_{\rm d} = L_{43}\cdot10^{43}$\,\unitLum\ is the accretion disk luminosity and $B = B_{-3}\cdot 10^{-3}$\,G. The external radiation field reprocesses accretion disk radiation with an optical depth of $\tau_{\rm ext} = 1\tau$ at a distance of $R_{\rm ext} = R_{16}\cdot 10^{16}$\,cm from the central engine. Thus, at relatively large viewing angles, \,TeV $\gamma$-rays are fewer when $B$ is weak. This effect becomes more pronounced as primary $\gamma$-rays are injected close to the central engine, or in dense BLR radiation fields. Producing $\gamma$-rays ever-closer to the BLR's outer edge (i.e. at greater $z^\gamma_0$) reduces their likelihood of suffering a substantial amount of $\gamma-\gamma$ collisions and losing their energy before escaping the AGN environment. Thus, \,TeV $\gamma$-rays may still be observed at low B, provided the primaries are produced close enough to the outer edge.

\subsection{BLR energy density}
\label{sec:BLRdensity}
\begin{figure}
    \centering
    \includegraphics[width=0.95\linewidth]{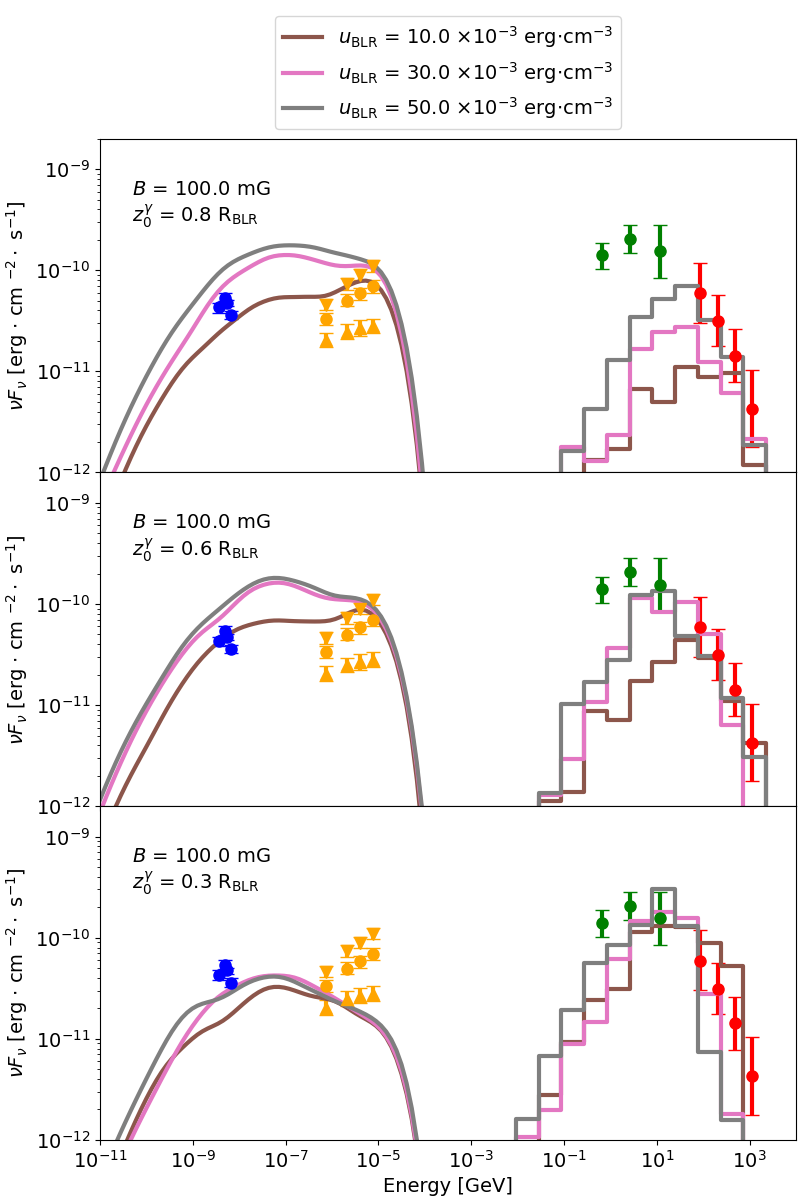}
    \caption{Cascade SEDs of simulations run at different BLR energy densities. The magnetic field was $B = 100$\,mG and primary photons were injected at a height of 0.8\,\unitRBLR, 0.6\,\unitRBLR\ and 0.3\,\unitRBLR\ in the \emph{upper}, \emph{centre} and \emph{lower} panels, respectively. The data points are as described in the caption of Figure~\ref{fig:SED_Bu50z0.8_varB_mu0.94-0.98}.
    }
    \label{fig:SED_B100uz0.8_varu_mu0.94-0.98}
\end{figure}

This subsection explores the effect of the BLR energy density on the cascade SEDs. We fix the primary photon injection height to 0.8\,\unitRBLR, 0.6\,\unitRBLR\ and 0.3\,\unitRBLR. To fully appreciate its effect across the entire spectrum, for reasons discussed in Section~\ref{sec:Bfield}, the magnetic field is fixed to 100\,mG. In general, for a given $(B, z^\gamma_0)$ combination, IC and synchrotron fluxes increase with increasing BLR energy density as shown by Figure~\ref{fig:SED_B100uz0.8_varu_mu0.94-0.98}. In an external radiation field of photon number density $n_{\rm ph}(\epsilon)$ and energy density $u_{\rm ext}(\epsilon)$, its absorption coefficient to a $\gamma$-ray of energy $\epsilon_\gamma$ in the delta-function approximation of the $\gamma\gamma$ cross-section \citep{dermer2009high} is
\begin{equation}
    \kappa_{\gamma\gamma} \approx \frac{2}{3}\frac{\sigma_T}{\epsilon_\gamma} n_{\rm ph}(2/\epsilon_\gamma) \sim \frac{\sigma_T}{3m_{\rm e}c^2}u_{\rm ext}(2/\epsilon_\gamma).
    \label{eq:kgg_approx}
\end{equation}
From expression~\eqref{eq:kgg_approx} one can intuitively understand that an increase in $u_{\rm BLR}$ leads to an increase in the development of cascades, and hence the cascade flux. However, the situation is more complicated for synchrotron emission: its luminosity is also in competition with that of the BLR. Suppose a region of radius $R_{\rm rad}$ is permeated by a radiation field of energy density $u_{\rm rad}$. Then its luminosity in the region is
\begin{equation}
    L_{\rm rad} \sim 4\pi cR^2_{\rm rad} u_{\rm rad}.
    \label{eq:Luminosity}
\end{equation}

Given that the IC luminosity $L_{\rm IC}$ is proportional to the target radiation field, comparing the luminosity of synchrotron emission, $L_{\rm sy}$, to $L_{\rm IC}$, in the AGN volume of radius $R_{\rm BLR}$, gives $L_{\rm sy}/L_{\rm IC} = u_{B}/u_{\rm rad} \propto u_{B}/u_{\rm BLR}$, where $u_B$ is the magnetic field energy density. While cascade development contributes to $L_{\rm sy}$, there is a limit to the cascade synchrotron emission's ability to reproduce the observed optical/UV -- X-ray emission, as it becomes throttled by this ratio with further increases to $u_{\rm BLR}$. Our simulation results confirm this trend. The cascade synchrotron rise-to-suppression transition is seen in the centre panel of Figure~\ref{fig:SED_B100uz0.8_varu_mu0.94-0.98}. In this panel, decreasing $z^\gamma_0$ from 0.8\,\unitRBLR\ to 0.6\,\unitRBLR\ caused a rise in cascade synchrotron emission. SEDs corresponding to $u_{\rm BLR} = 30\times 10^{-3}$\,\unitU\ and $50\times 10^{-3}$\,\unitU\, at this intermediate height start showing synchrotron saturation in optical/UV to X-ray emissions. In the lower panel of this figure, $z^\gamma_0$ is $0.3\,\unitRBLR$ and optical/UV -- soft X-ray emission from a $u_{\rm BLR}$ of $10\times 10^{-3}\,\unitU$ continues to rise, while the optical/UV emissions from $u_{\rm BLR} = (30\rm\ and\ 50)\,\times 10^{-3}\,\unitU$ have reached saturation and their X-ray emission experiences suppression. Lowering $z^\gamma_0$ mimics the effect of increasing $u_{\rm BLR}$; this is discussed in depth in Section~\ref{sec:R0}. While the cascade simulations demonstrate an external-radiation-field-driven evolution characterised by a rise and fall in cascade synchrotron emission, the cascade IC emission also exhibits a kind of rise and fall behaviour in response to $u_{\rm BLR}$. Its rise with $u_{\rm BLR}$ is simply due to an increase in cascade development, as stated earlier. However, as $u_{\rm BLR}$ increases, so does the optical depth to energetic $\gamma$-ray photons. We use the lower panel of Figure~\ref{fig:SED_B100uz0.8_varu_mu0.94-0.98} to best illustrate this effect. At $z^\gamma_0 = 0.3$\,\unitRBLR, further increases in $u_{\rm BLR}$ results in the BLR eventually becoming optically thick to \,TeV photons.

\subsection{Height above the accretion disk}
\label{sec:R0}
\begin{figure}
    \centering
    \includegraphics[width=0.95\linewidth]{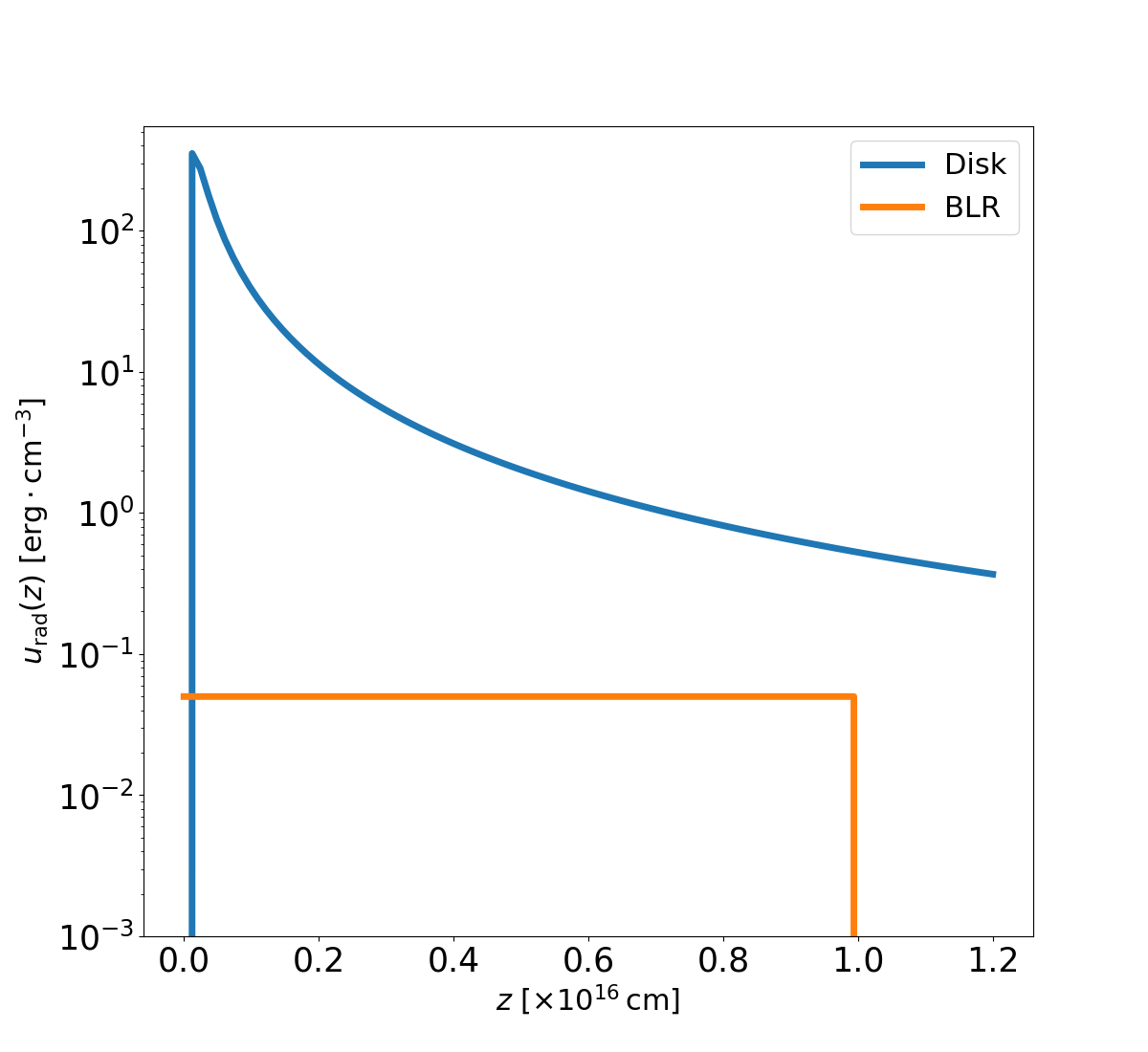}
    \caption{Energy densities of BLR and accretion disk radiation fields, respectively, with varying height, $z$, above the accretion disk. The disk luminosity is $L_{\rm d} = 1.88\times 10^{43}\,\unitLum$.}
    \label{fig:energydensity}
\end{figure}

We calculate the accretion disk energy density as a function of height above one side of the disk by integrating Equation~\eqref{eq:disk} over energy and the area of the accretion disk. This gives
\begin{align}
     u_{\rm d}(z) &= 
     \frac{m_ec^2}{2} 
         \int_0^{2\pi}\,d\phi_{\rm d}
         \int_{R_{in}}^{R_{out}} R_{\rm d}\,dR_{\rm d} 
         \int_0^{\infty} \epsilon\frac{d^2 n_{\rm ph}}{dA_{\rm d}d\epsilon}\,d\epsilon \nonumber\\
    &= 
    \frac{3}{2} \frac{\sigma_{SB}}{c}\left(\frac{m_ec^2}{k_B}\right)^4 |z|
        \int_{R_{in}}^{R_{out}} \frac{R_{\rm d} \Theta^4(R_{\rm d})}{(R_{\rm d}^2 + z^2)^{3/2}} \,dR_{\rm d}.
\label{eq:ud_z}
\end{align}
The energy density profiles of the BLR and SS disk radiation fields are plotted in Figure~\ref{fig:energydensity}.
\begin{figure}
    \centering
    \includegraphics[width=0.95\linewidth]{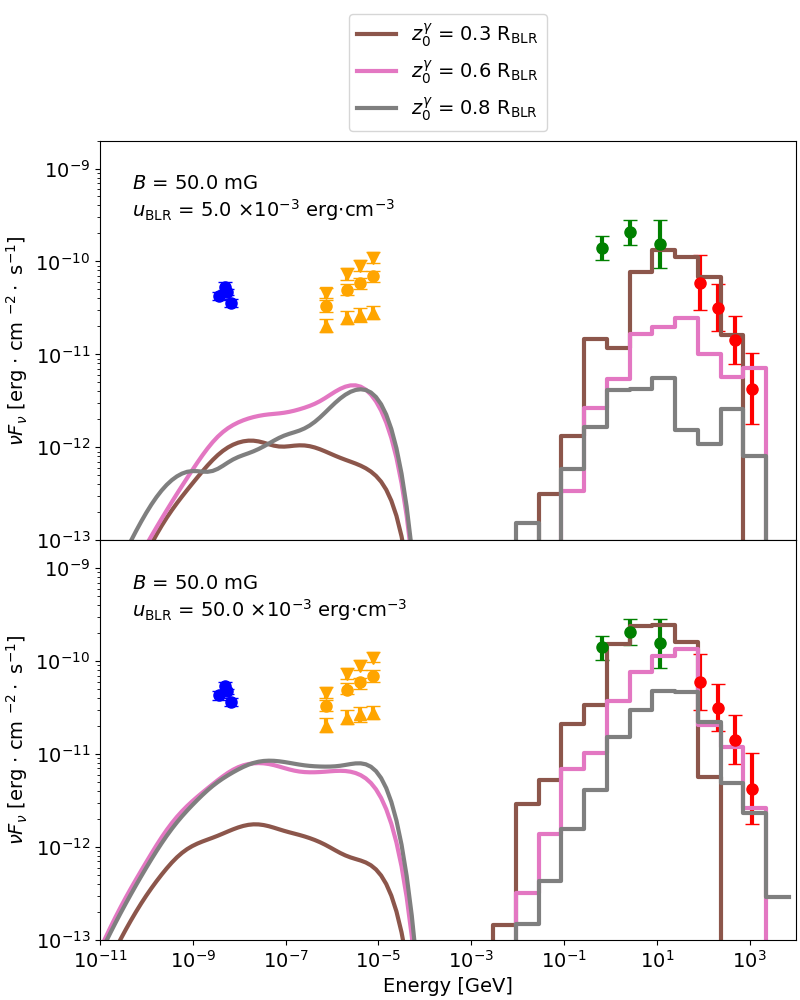}
    \caption{Cascade SEDs generated by changing injection heights of primary $\gamma$-rays above the accretion disk for fixed $B = 50$\,mG and $u_{\rm BLR} = 5\times 10^{-3}$ and $50\times 10^{-3}$\,\unitU\ in the \emph{upper} and \emph{lower} panels, respectively. The data points are as described in the caption of Figure~\ref{fig:SED_Bu50z0.8_varB_mu0.94-0.98}.
    }
    \label{fig:SED_B50u5zvarz_0.94-0.98}
\end{figure}

The profile of the SS disk energy density clearly illustrates the growing importance of the disk radiation field as the height $z$ decreases. We thus investigate the effect of varying the injection height, $z^\gamma_0$, of primary $\gamma$-rays, as it has a direct link to $u_{\rm d}$ and will show the interplay between the considered external radiation fields, $u_{\rm BLR}$ and $u_{\rm d}$. In these simulations, the magnetic field was fixed to 50\,mG and $u_{\rm BLR}$ to $5\times 10^{-3}$ and $50\times 10^{-3}$\,\unitU\ in the upper and lower panels of Figure~\ref{fig:SED_B50u5zvarz_0.94-0.98}, respectively. To emphasize the effect of $u_{\rm d}$, we minimized the contribution of $u_{\rm BLR}$ by simulating the weakest value of $5\times 10^{-3}$\,\unitU\ used in our previous simulations in the upper panel of Figure~\ref{fig:SED_B50u5zvarz_0.94-0.98}, where it is demonstrably clear that the IC cascade emission rises with decreasing $z^\gamma_0$. In this case, $u_{\rm d}$ intensifies closer to the central engine, increasing cascade development and thus IC emissions. The cascade SED corresponding to $z^\gamma_0 = 0.8$\,\unitRBLR\ represents the weakest $u_{\rm d}$. An evaluation of the cascade synchrotron emission shows this height to result in the weakest synchrotron flux, rising with decreasing height. However, just as in Section~\ref{sec:BLRdensity}, the ratio discussed in that section is reflected here, where a transition had started to take place at the heights $z^\gamma_0 = 0.6\,\unitRBLR\ \rm and\ 0.8$\,\unitRBLR, as seen in the lower panel of Figure~\ref{fig:SED_B50u5zvarz_0.94-0.98}, where $u_{\rm BLR} = 50\times 10^{-3}$\,\unitU. At this transitional height, optical/UV -- X-ray emissions have saturated at a stable level
\footnote{This is not a true stable level in an absolute sense. Dense external radiation enhances the efficiency of IC scattering, which extracts large amounts of energy from e$^\pm$ and leaving them with less energy available for synchrotron emission.
}. In the upper panel, injection heights 0.6\,\unitRBLR\ and 0.8\,\unitRBLR\ continue to raise their overall optical/UV -- X-ray emissions. Increasing $u_{\rm BLR}$, the lower panel shows the 0.6\,\unitRBLR\ height level starting to show signs of X-ray suppression. Thus a denser BLR means the transition can occur further from the central engine. This figure seems to suggest that further decreasing $z^\gamma_0$ has the same effect discussed in Section~\ref{sec:BLRdensity} of further increasing $u_{\rm BLR}$. We point out, however that, in general, the statement holds for the density of any external optical/UV photon field, since e$^\pm$ pairs do not distinguish their origin, whether from the accretion disk or the BLR. The proximity of the primary $\gamma$-rays to the outer edge of the BLR reduces their likelihood of experiencing $\gamma-\gamma$ absorption, making for an easier escape of photons of all energies as seen in the upper panel of Figure~\ref{fig:SED_B50u5zvarz_0.94-0.98}. However, the combined effects of increasing $u_{\rm BLR}$ and decreasing $z^\gamma_0$ compound on the optical depth of the highest energy photons. In the lower panel of Figure~\ref{fig:SED_B50u5zvarz_0.94-0.98} we observe that at $z^\gamma_0 = 0.3$\,\unitRBLR, \,TeV $\gamma$-rays are significantly absorbed as compared to the same injection height in the upper panel of this figure, where $u_{\rm BLR}$ is weaker.

\subsection{Comparison  with the SED of NGC 1275}
\label{sec:SEDcontribution}
The parameter study explored in the previous three subsections gave insights into the AGN broadband SED's behaviour to $B, u_{\rm BLR}$ and $z^\gamma_0$. Based on these, we use a possible ($B, u_{\rm BLR}, z_0^\gamma$)-combination of parameters that can plausibly reproduce the SED of NGC 1275 during flares similar to that of the night of 21 December 2022. Figures~\ref{fig:SED_B100uz0.8_varu_mu0.94-0.98} and~\ref{fig:SED_B50u5zvarz_0.94-0.98} will play a central role towards discussions in this endeavour. The magnetic field certainly plays a vital role in the flux normalisation in both synchrotron and IC components. Also very importantly, it influences the shape of the IC spectrum. Provided the AGN environment is permeated by an oblique magnetic field with a modest transverse component, these results clearly suggest $B$-field strengths of multi-tens to hundreds of \,mG are sufficient to explain NGC 1275's broadband SED. On the night of 21 December 2022, the MACE detected a \,TeV flare from NGC 1275. In \citet{godambe2024very}, the authors constructed its multi-wavelength SED with quasi-simultaneous data from the \emph{Fermi}-LAT and \emph{Swift}-UVOT. They collected X-ray data from the \emph{Swift}-XRT from MJDs 59933.9 and 59935.9, which correspond to 20 and 22 December 2022. They thus use as X-ray flux representative of 21 December 2022 the average X-ray flux of these two days. Figure~\ref{fig:SED_B50u5zvarz_0.94-0.98} shows that simulations performed at $B=50$\,mG still underestimate the optical/UV -- X-ray flux activity of MJD 59933.9. To reach X-ray flux levels on MJD 59935.9, a comparison of Figure~\ref{fig:SED_B100uz0.8_varu_mu0.94-0.98} and~\ref{fig:SED_B50u5zvarz_0.94-0.98} suggests that a magnetic field strength of about twofold more was required. In this section, we attempt to model the flaring episode of 21 December 2022, with these two overlapping days constraining the magnetic field.

Beginning at the deepest simulated injection region, $z^\gamma_0 = 0.3$\,\unitRBLR, where $u_{\rm d}$ plays a significant role, Figure~\ref{fig:SED_B50u5zvarz_0.94-0.98} shows that the BLR energy density needs to be at least few tens of $\times 10^{-3}$\,\unitU\ to reproduce \emph{Fermi} emissions. However, cascades initiated by primary $\gamma$-rays injected deep within the AGN are severely attenuated in the MACE energy range at both $B=50$\,mG and $100$\,mG. Magnetic fields stronger than $\sim 100$\,mG would overestimate the high X-ray state of MJD 59935.9. Figures~\ref{fig:SED_B100uz0.8_varu_mu0.94-0.98} and \ref{fig:SED_B50u5zvarz_0.94-0.98} seem to suggest that primary $\gamma$-rays that initiate IC-supported cascades at regions further from the central engine lead to observable VHE emissions. Furthermore, \emph{Fermi}-LAT emissions are produced by IC-supported cascades initiated close to the central engine. While the sources of optical/UV and X-ray emissions seem to be located further away from the central engine, this conclusion is still tentative, given the limitation placed by the upper limits at these frequencies. The X-ray fluxes on MJDs 59933.9 and 59935.9 constrain the magnetic field to be in the range between $\sim 50$\,mG and $\sim 100$\,mG.

\begin{figure}
    \centering
    \includegraphics[width=0.99\linewidth]{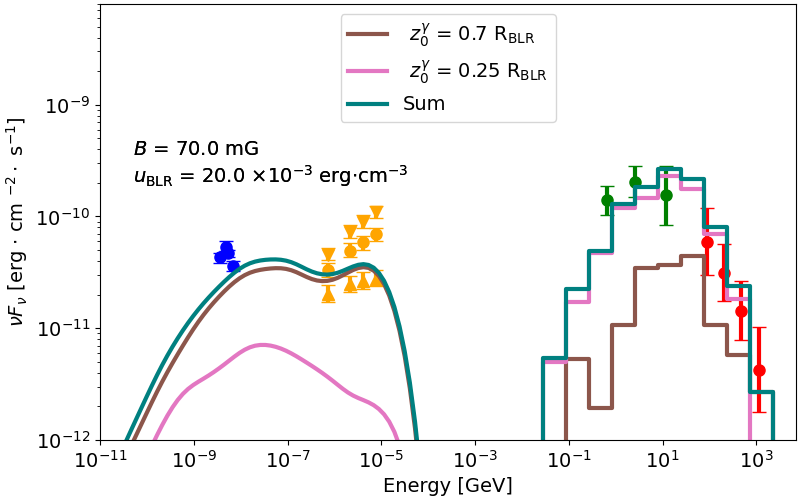}
    \caption{Cascade contribution fitted to NGC 1275's SED during a flaring episode of 21 December 2022. For consistency, in the X-ray, the average X-ray flux of MJDs 59933.9 and 59935.9, corresponding to 20 and 22 December 2022, respectively, which is representative of 21 December 2022 is fitted. \,TeV $\gamma$-rays were reproduced by cascades initiated by primary $\gamma$-rays produced at intermediate heights to the BLR's outer edge. While the \emph{Fermi}-LAT emissions are from cascades initiated at lower heights. Optical/UV and X-ray cannot be properly accounted for by the cascades since the flux data in these frequencies may be contaminated by emissions by the host galaxy. The fit values corresponding to the constant $B$ and $u_{\rm BLR}$ are shown in the figure, with the injection height values described in the legend, namely \emph{pink} being the lower height of 0.25\,\unitRBLR\ and \emph{brown} being the intermediate to larger height of 0.7\,\unitRBLR. Our SED fit from the cascades is represented by the teal line, which is the sum of these two SEDs. The data points are as described in the caption of Figure~\ref{fig:SED_Bu50z0.8_varB_mu0.94-0.98}, in the case of X-ray data the model fit is to the \emph{round} dots.
    }
    \label{fig:SEDcontrib_B100uz0.3_varu_mu0.94-0.98}
\end{figure}

In Figure~\ref{fig:SEDcontrib_B100uz0.3_varu_mu0.94-0.98}, we fit the quasi-simultaneous broadband SED of NGC 1275 using the above considerations. These quasi-simultaneous data, in line with the MACE detection, correspond to 21 December 2022; therefore in the case of X-ray data, we focus our fits to the average X-ray flux, i.e. the round dots. The corresponding fit parameters are shown in the figure, with the five fixed parameters as described in Section~\ref{sec:codesetup}. From the parameter combinations currently simulated, emissions from the cascades show potential to reproduce the broadband SED. We used the sum of cascade SEDs, with similar input $B$ and $u_{\rm BLR}$, and different injection heights of $z_0^\gamma = 0.25\,\unitRBLR$ to produce \emph{Fermi}-LAT emissions and $0.7\,\unitRBLR$ to produce MACE emissions. The MACE spectrum is reproduced reasonably well. With the varied three parameters, the cascade code can partially reproduce the \emph{Fermi}-LAT spectrum. For this, a slightly larger $\theta_{B}$ may be required. However, this spectral discrepency may also be an indication that, in addition to the external Compton scattering of BLR and accretion disk photons, the cascade code may require a non-negligible component from the Comptonization of the cascade optical/UV and X-ray emissions. 

\section{Summary and conclusion}
\label{sec:discussion}
NGC 1275 is currently accepted as a type 1.5 Seyfert radio galaxy. However, a review of the literature suggests its classification is still not fully understood. Its classification seems to depend on the time of observing campaigns and the data quality. Some studies classified it as a type 2, while others could not reach a definite conclusion \citep[see][and references within]{punsly2018revealing, marin2025spectropolarimetry}. Adding to the complexity of AGNi is the existence of repeating changing-look AGNi (RCL AGNi) \citep{dong2025discovery}. Studies have not confirmed NGC 1275 to be an RCL AGN. However, with its behaviour exhibiting a mixture of features from these AGN types, it may be reasonable to expect it to be a RCL AGN. Guided by this justification, it therefore seemed reasonable to model NGC 1275 with the presence of a BLR.

We used the cascade code with external Comptonization of the BLR and SS accretion disk radiation fields as a mechanism to explain the HE to VHE $\gamma$-ray emissions from a flare event of NGC 1275. The cascade code does not rely on Doppler beaming of the emission from AGNi. We investigated the effect of three out of the eight parameters required by the code. The fixed values, guided by the literature, of five of these parameters were listed in Section~\ref{sec:codesetup}. Three of them are still largely empirically unknown, and left as free parameters and thus we began an initial parameter study with them. While synchrotron behaviour was typical in response to the magnetic field, the effect was rather suprising in the IC component. Firstly, the component of the magnetic field transverse to the initial direction of the primary $\gamma$-rays was very efficient in the isotropization of the e$^\pm$ pairs. This is very important in explaining $\gamma$-ray emissions at large angles w.r.t the jet-axis. Secondly, the strength of the $B$ field seems to regulate the emission band of the $\gamma$-ray spectrum. The Larmor radius of the pairs may play a role in this behaviour, the combination of the high energy of the e$^\pm$ and low $B$ field increases $r_L$ resulting in less efficient deflection of the particles, such that high $\gamma_{\rm e^\pm}$ particles mostly radiate in the forward jet direction, and thus their emission is not detected at the large viewing angles typical of radio galaxies.
Cascade emissions are largely azimuthaly symmetric, with azimuthal symmetry mildly broken by the orientation of an oblique magnetic field.

By varying $u_{\rm BLR}$ and $z^\gamma_0$, our simulations revealed the expected interplay between the BLR and accretion disk radiation fields. The SED reaction to increasing $u_{\rm BLR}$ is similar to decreasing $z^\gamma_0$, as is indicated by the plot of $u_{\rm BLR}$ and $u_{\rm d}$ vs height above the accretion disk plane (see Figure~\ref{fig:energydensity}). Since a decrease in $z^\gamma_0$ corresponds to an increase in $u_{\rm d}$, we collectively discuss effects of $u_{\rm BLR}$ and $z^\gamma_0$ by referring to the \emph{external radiation field energy density}, unless a comment is made explicitly about $u_{\rm BLR}$ or $z^\gamma_0$ in text. A dense external radiation field naturally leads to an increase in the cascade development. This was reflected in the SED results in Sections~\ref{sec:BLRdensity} and \ref{sec:R0}, where both synchrotron and IC fluxes increased with increasing $u_{\rm BLR}$ and decreasing $z^\gamma_0$, respectively. However, in the synchrotron component there is limit to its progressive increase. Eventually, Compton cooling dominates over synchrotron losses, thus suppressing the synchrotron luminosity, initially in X-rays. In the IC component we also observe a flux increase with external radiation energy density. However, further increases to the external radiation energy density, increases its optical depth to energetic photons, which truncates the highest energy $\gamma$-rays.

The limitations placed by the external radiation fields have implications on possible origins of the primary $\gamma$-rays. Those produced deep within the central engine
better explain the \emph{Fermi}-LAT spectrum, while the MACE spectrum is better explained by $\gamma$-rays produced at intermediate heights to the BLR's outer edge. These different $\gamma$-ray injection sites suggested by the cascade code to explain various spectra support multi-zone emission models.

Although this is not in the scope of the current work, we briefly comment on the cascade variability timescales. During NGC1275's reported flare \citep{godambe2024very}, the \emph{Fermi} light curve showed variability on an average timescale of $\sim 4.3$ days. The cascade variability timescale of \emph{Fermi} photons ranged from few hours ($< 3$\,hours) at the strongest external radiation energy densities to about $\sim 30$ minutes at the weakest simulated values. Notably, this is much shorter than the light-crossing time of the BLR, $R_{\rm BLR} = 10^{16}$\,cm, which is $R_{\rm BLR}/c\sim 4$\,days. Thus, the observed $\sim 4.3$\,day variability suggests that the cascade variability is smeared out by the source's intrinsic variability.

\section{Future outlook}
\label{sec:future}
This work demonstrates that inverse-Compton-supported cascades have the potential to explain multi-wavelength emissions detected in radio galaxies, without the need for Doppler beaming. Spectropolarimetry studies revealed helical magnetic field structures in blazars \citep{andati2024spectropolarimetric}. The simulations showed that cascade synchrotron energy densities are not negligible. Together with \emph{Fermi}-LAT spectral data, this combination may be contending that the cascade code is missing the synchrotron self-Compton (SSC) component of these synchrotron seed photons. In future works, we will extend the cascade code to account for SSC emission from cascade e$^\pm$ pairs and start with a toroidal magnetic field, a simplified version of the helical $B$ field. Also, while the parameter study explored in the current work explained the cascade flux levels in the SEDs, as a future prospect, we will explore other parameter variations and examine their effect on the IC spectral shape and variability.

\section*{Acknowledgements}

We want to thank the National Research Foundation (NRF) of South Africa for their financial support through SA-GAMMA, BRICS STI and NITheCS grants, and through a PhD bursary to M.N. We are also grateful for the opportunities to use the high-performance computing facility at the University of Johannesburg and the National Integrated CyberInfrastructure System.
Without these resources, this research would not have been possible.

\section*{Data Availability}
Fitted data presented here was taken from \citet{godambe2024very}.



\bibliographystyle{mnras}
\bibliography{./refs_mfp}

\begin{thebibliography}{}
\makeatletter
\relax
\def\mn@urlcharsother{\let\do\@makeother \do\$\do\&\do\#\do\^\do\_\do\%\do\~}
\def\mn@doi{\begingroup\mn@urlcharsother \@ifnextchar [ {\mn@doi@}
  {\mn@doi@[]}}
\def\mn@doi@[#1]#2{\def\@tempa{#1}\ifx\@tempa\@empty \href
  {http://dx.doi.org/#2} {doi:#2}\else \href {http://dx.doi.org/#2} {#1}\fi
  \endgroup}
\def\mn@eprint#1#2{\mn@eprint@#1:#2::\@nil}
\def\mn@eprint@arXiv#1{\href {http://arxiv.org/abs/#1} {{\tt arXiv:#1}}}
\def\mn@eprint@dblp#1{\href {http://dblp.uni-trier.de/rec/bibtex/#1.xml}
  {dblp:#1}}
\def\mn@eprint@#1:#2:#3:#4\@nil{\def\@tempa {#1}\def\@tempb {#2}\def\@tempc
  {#3}\ifx \@tempc \@empty \let \@tempc \@tempb \let \@tempb \@tempa \fi \ifx
  \@tempb \@empty \def\@tempb {arXiv}\fi \@ifundefined
  {mn@eprint@\@tempb}{\@tempb:\@tempc}{\expandafter \expandafter \csname
  mn@eprint@\@tempb\endcsname \expandafter{\@tempc}}}

\bibitem[\protect\citeauthoryear{{Abdo} et~al.,}{{Abdo}
  et~al.}{2009}]{abdo2009Fermi}
{Abdo} A.~A.,  et~al., 2009, \mn@doi [\apj] {10.1088/0004-637X/707/2/1310},
  \href {https://ui.adsabs.harvard.edu/abs/2009ApJ...707.1310A} {707, 1310}

\bibitem[\protect\citeauthoryear{{Abdo} et~al.,}{{Abdo}
  et~al.}{2010}]{abdo2010fermi}
{Abdo} A.~A.,  et~al., 2010, \mn@doi [\apj] {10.1088/0004-637X/723/2/1082},
  \href {https://ui.adsabs.harvard.edu/abs/2010ApJ...723.1082A} {723, 1082}

\bibitem[\protect\citeauthoryear{{Ajello} et~al.,}{{Ajello}
  et~al.}{2020}]{ajello2020fourth}
{Ajello} M.,  et~al., 2020, \mn@doi [\apj] {10.3847/1538-4357/ab791e}, 892, 105

\bibitem[\protect\citeauthoryear{{Andati}, {Baidoo}, {Ramaila}, {Smirnov},
  {Makhathini}  \& {Perley}}{{Andati}
  et~al.}{2024}]{andati2024spectropolarimetric}
{Andati} L. A.~L.,  {Baidoo} L.~M.,  {Ramaila} A. J.~T.,  {Smirnov} O.~M.,
  {Makhathini} S.,   {Perley} R.~A.,  2024, \mn@doi [\mnras]
  {10.1093/mnras/stae598}, \href
  {https://ui.adsabs.harvard.edu/abs/2024MNRAS.529.1626A} {529, 1626}

\bibitem[\protect\citeauthoryear{{Asada}, {Kameno}, {Shen}, {Horiuchi},
  {Gabuzda}  \& {Inoue}}{{Asada} et~al.}{2006}]{asada2006expanding}
{Asada} K.,  {Kameno} S.,  {Shen} Z.-Q.,  {Horiuchi} S.,  {Gabuzda} D.~C.,
  {Inoue} M.,  2006, \mn@doi [\pasj] {10.1093/pasj/58.2.261}, \href
  {https://ui.adsabs.harvard.edu/abs/2006PASJ...58..261A} {58, 261}

\bibitem[\protect\citeauthoryear{{Blandford}, {Meier}  \&
  {Readhead}}{{Blandford} et~al.}{2019}]{blandford2019relativistic}
{Blandford} R.,  {Meier} D.,   {Readhead} A.,  2019, \mn@doi [\araa]
  {https://doi.org/10.1146/annurev-astro-081817-051948}, 57, 467

\bibitem[\protect\citeauthoryear{{B{\"o}ttcher} \&
  {Schlickeiser}}{{B{\"o}ttcher} \& {Schlickeiser}}{1997}]{bottcher1997pair}
{B{\"o}ttcher} M.,  {Schlickeiser} R.,  1997, \mn@doi [\aap]
  {10.48550/arXiv.astro-ph/9703069}, \href
  {https://ui.adsabs.harvard.edu/abs/1997A&A...325..866B} {325, 866}

\bibitem[\protect\citeauthoryear{{B{\"o}ttcher}, {Mause}  \&
  {Schlickeiser}}{{B{\"o}ttcher} et~al.}{1997}]{bottcher1997gammaray}
{B{\"o}ttcher} M.,  {Mause} H.,   {Schlickeiser} R.,  1997, A\&A, \href
  {https://ui.adsabs.harvard.edu/abs/1997A&A...324..395B} {324, 395}

\bibitem[\protect\citeauthoryear{Calistro~{Rivera}, {Lusso}, {Hennawi}  \&
  {Hogg}}{Calistro~{Rivera} et~al.}{2016}]{rivera2016agnfitter}
Calistro~{Rivera} G.,  {Lusso} E.,  {Hennawi} J.~F.,   {Hogg} D.~W.,  2016,
  \mn@doi [ApJ] {10.3847/1538-4357/833/1/98}, 833, 98

\bibitem[\protect\citeauthoryear{{Cao} et~al.,}{{Cao}
  et~al.}{2024}]{cao2024detection}
{Cao} Z.,  et~al., 2024, \mn@doi [\mnras] {10.1093/mnras/stae2512}, \href
  {https://ui.adsabs.harvard.edu/abs/2024MNRAS.tmp.2493C} {p. stae2512}

\bibitem[\protect\citeauthoryear{{Dermer} \& {Giebels}}{{Dermer} \&
  {Giebels}}{2016}]{dermer2016active}
{Dermer} C.~D.,  {Giebels} B.,  2016, \mn@doi [Comptes Rendus Physique]
  {https://doi.org/10.1016/j.crhy.2016.04.004}, 17, 594

\bibitem[\protect\citeauthoryear{{Dermer} \& {Menon}}{{Dermer} \&
  {Menon}}{2009}]{dermer2009high}
{Dermer} C.~D.,  {Menon} G.,  2009, High Energy Radiation from Black Holes:
  Gamma Rays, Cosmic Rays, and Neutrinos

\bibitem[\protect\citeauthoryear{{Dong}, {Zhang}, {Gu}, {Sun}, {Guo}, {Cai},
  {Wang}  \& {Zheng}}{{Dong} et~al.}{2025}]{dong2025discovery}
{Dong} Q.,  {Zhang} Z.-X.,  {Gu} W.-M.,  {Sun} M.,  {Guo} W.-J.,  {Cai} Z.-Y.,
  {Wang} J.-X.,   {Zheng} Y.-G.,  2025, \mn@doi [arXiv e-prints]
  {10.48550/arXiv.2510.18445}, \href
  {https://ui.adsabs.harvard.edu/abs/2025arXiv251018445D} {p. arXiv:2510.18445}

\bibitem[\protect\citeauthoryear{{Ghisellini} \& {Tavecchio}}{{Ghisellini} \&
  {Tavecchio}}{2009}]{ghisellini2009canonical}
{Ghisellini} G.,  {Tavecchio} F.,  2009, MNRAS, 397, 985

\bibitem[\protect\citeauthoryear{{Godambe} et~al.,}{{Godambe}
  et~al.}{2024}]{godambe2024very}
{Godambe} S.,  et~al., 2024, \mn@doi [\apjl] {10.3847/2041-8213/ad8083}, \href
  {https://ui.adsabs.harvard.edu/abs/2024ApJ...974L..31G} {974, L31}

\bibitem[\protect\citeauthoryear{{Jauch} \& {Rohrlich}}{{Jauch} \&
  {Rohrlich}}{1976}]{jauch1976theory}
{Jauch} J.~M.,  {Rohrlich} F.,  1976, Theory of the External Field.
Springer Berlin Heidelberg, Berlin, Heidelberg, pp 302--326,
  \mn@doi{10.1007/978-3-642-80951-4_14}, \url
  {https://doi.org/10.1007/978-3-642-80951-4_14}

\bibitem[\protect\citeauthoryear{{Krichbaum} et~al.,}{{Krichbaum}
  et~al.}{1992}]{krichbaum1992evolution}
{Krichbaum} T.~P.,  et~al., 1992, \aap, \href
  {https://ui.adsabs.harvard.edu/abs/1992A&A...260...33K} {260, 33}

\bibitem[\protect\citeauthoryear{{Marin}, {Pursimo}, {Liodakis}, {Lindfors},
  {Biedermann}, {Hutsem{\'e}kers}  \& {Turkki}}{{Marin}
  et~al.}{2025}]{marin2025spectropolarimetry}
{Marin} F.,  {Pursimo} T.,  {Liodakis} I.,  {Lindfors} E.,  {Biedermann} J.,
  {Hutsem{\'e}kers} D.,   {Turkki} M.,  2025, \mn@doi [\aap]
  {10.1051/0004-6361/202556163}, \href
  {https://ui.adsabs.harvard.edu/abs/2025A&A...702L..16M} {702, L16}

\bibitem[\protect\citeauthoryear{{Martí}}{{Martí}}{2019}]{marti2019numerical}
{Martí} J.-M.,  2019, \mn@doi [Galaxies] {10.3390/galaxies7010024}, 7

\bibitem[\protect\citeauthoryear{{Netzer} \& {Laor}}{{Netzer} \&
  {Laor}}{1993}]{netzer1993dust}
{Netzer} H.,  {Laor} A.,  1993, \mn@doi [\apjl] {10.1086/186741}, \href
  {https://ui.adsabs.harvard.edu/abs/1993ApJ...404L..51N} {404, L51}

\bibitem[\protect\citeauthoryear{{Ntshatsha}, {B{\"o}ttcher}  \&
  {Razzaque}}{{Ntshatsha} et~al.}{2024}]{ntshatsha2024compton}
{Ntshatsha} M.,  {B{\"o}ttcher} M.,   {Razzaque} S.,  2024, in High Energy
  Astrophysics in Southern Africa (HEASA2023). , \mn@doi{10.22323/1.459.0014},
  \url {https://hdl.handle.net/10210/512008}

\bibitem[\protect\citeauthoryear{{Punsly}, {Marziani}, {Bennert}, {Nagai}  \&
  {Gurwell}}{{Punsly} et~al.}{2018}]{punsly2018revealing}
{Punsly} B.,  {Marziani} P.,  {Bennert} V.~N.,  {Nagai} H.,   {Gurwell} M.~A.,
  2018, \mn@doi [\apj] {10.3847/1538-4357/aaec75}, 869, 143

\bibitem[\protect\citeauthoryear{{Pushkarev}, {Kovalev}, {Lister}  \&
  {Savolainen}}{{Pushkarev} et~al.}{2009}]{pushkarev2009jet}
{Pushkarev} A.~B.,  {Kovalev} Y.~Y.,  {Lister} M.~L.,   {Savolainen} T.,  2009,
  \mn@doi [A&A] {10.1051/0004-6361/200913422}, 507, L33

\bibitem[\protect\citeauthoryear{{Reynolds} et~al.,}{{Reynolds}
  et~al.}{2021}]{reynolds2021probing}
{Reynolds} C.~S.,  et~al., 2021, \mn@doi [\mnras] {10.1093/mnras/stab2507},
  507, 5613

\bibitem[\protect\citeauthoryear{{Robson}}{{Robson}}{1996}]{robson1996active}
{Robson} I.,  1996, Active Galactic Nuclei.
John Wiley \& Sons Ltd

\bibitem[\protect\citeauthoryear{{Roustazadeh} \& {B{\"o}ttcher}}{{Roustazadeh}
  \& {B{\"o}ttcher}}{2010}]{roustazadeh2010very}
{Roustazadeh} P.,  {B{\"o}ttcher} M.,  2010, \mn@doi [\apj]
  {10.1088/0004-637X/717/1/468}, \href
  {https://ui.adsabs.harvard.edu/abs/2010ApJ...717..468R} {717, 468}

\bibitem[\protect\citeauthoryear{{Roustazadeh} \& {B{\"o}ttcher}}{{Roustazadeh}
  \& {B{\"o}ttcher}}{2011}]{roustazadeh2011very}
{Roustazadeh} P.,  {B{\"o}ttcher} M.,  2011, \mn@doi [\apj]
  {10.1088/0004-637X/728/2/134}, \href
  {https://ui.adsabs.harvard.edu/abs/2011ApJ...728..134R} {728, 134}

\bibitem[\protect\citeauthoryear{{Roustazadeh} \& {B{\"o}ttcher}}{{Roustazadeh}
  \& {B{\"o}ttcher}}{2012}]{roustazadeh2012synchrotron}
{Roustazadeh} P.,  {B{\"o}ttcher} M.,  2012, \mn@doi [\apj]
  {10.1088/0004-637X/750/1/26}, \href
  {https://ui.adsabs.harvard.edu/abs/2012ApJ...750...26R} {750, 26}

\bibitem[\protect\citeauthoryear{{Roustazadeh}, {Thrush}  \&
  {B{\"o}ttcher}}{{Roustazadeh} et~al.}{2015}]{roustazadeh2015time}
{Roustazadeh} P.,  {Thrush} S.,   {B{\"o}ttcher} M.,  2015, in 3rd Annual
  Conference on High Energy Astrophysics in Southern Africa (HEASA2015). p.~18,
  \mn@doi{10.22323/1.241.0018}

\bibitem[\protect\citeauthoryear{{Rybicki} \& {Lightman}}{{Rybicki} \&
  {Lightman}}{1986}]{rybicki1996radiative}
{Rybicki} G.~B.,  {Lightman} A.~P.,  1986, Radiative processes in astrophysics.
John Wiley \& Sons

\bibitem[\protect\citeauthoryear{{Shakura} \& {Sunyaev}}{{Shakura} \&
  {Sunyaev}}{1973}]{shakura1973black}
{Shakura} N.~I.,  {Sunyaev} R.~A.,  1973, \aap, \href
  {https://ui.adsabs.harvard.edu/abs/1973A&A....24..337S} {24, 337}

\bibitem[\protect\citeauthoryear{{Sitarek} \& {Bednarek}}{{Sitarek} \&
  {Bednarek}}{2010}]{sitarek2010timedependent}
{Sitarek} J.,  {Bednarek} W.,  2010, \mn@doi [MNRAS]
  {10.1111/j.1365-2966.2010.17330.x}, \href
  {https://ui.adsabs.harvard.edu/abs/2010MNRAS.409..662S} {409, 662}

\bibitem[\protect\citeauthoryear{{Stern} \& {Poutanen}}{{Stern} \&
  {Poutanen}}{2014}]{stern2014mystery}
{Stern} B.~E.,  {Poutanen} J.,  2014, ApJ, 794, 8

\bibitem[\protect\citeauthoryear{{Tanada}, {Kataoka}, {Arimoto}, {Akita},
  {Cheung}, {Digel}  \& {Fukazawa}}{{Tanada} et~al.}{2018}]{tanada2018origins}
{Tanada} K.,  {Kataoka} J.,  {Arimoto} M.,  {Akita} M.,  {Cheung} C.~C.,
  {Digel} S.~W.,   {Fukazawa} Y.,  2018, \mn@doi [\apj]
  {10.3847/1538-4357/aac26b}, \href
  {https://ui.adsabs.harvard.edu/abs/2018ApJ...860...74T} {860, 74}

\bibitem[\protect\citeauthoryear{{Urry} \& {Padovani}}{{Urry} \&
  {Padovani}}{1995}]{urry1995unified}
{Urry} C.~M.,  {Padovani} P.,  1995, \mn@doi [\pasp] {10.1086/133630}, \href
  {https://ui.adsabs.harvard.edu/abs/1995PASP..107..803U} {107, 803}

\bibitem[\protect\citeauthoryear{{Vermeulen}, {Readhead}  \&
  {Backer}}{{Vermeulen} et~al.}{1994}]{vermeulen1994discovery}
{Vermeulen} R.~C.,  {Readhead} A.~C.~S.,   {Backer} D.~C.,  1994, \mn@doi
  [\apjl] {10.1086/187433}, \href
  {https://ui.adsabs.harvard.edu/abs/1994ApJ...430L..41V} {430, L41}

\bibitem[\protect\citeauthoryear{{Walker}, {Romney}  \& {Benson}}{{Walker}
  et~al.}{1994}]{walker1994detection}
{Walker} R.~C.,  {Romney} J.~D.,   {Benson} J.~M.,  1994, \mn@doi [\apjl]
  {10.1086/187434}, \href
  {https://ui.adsabs.harvard.edu/abs/1994ApJ...430L..45W} {430, L45}

\makeatother
\end{thebibliography}




\appendix
\section{Angular distribution}
This is a 2D histogram similar to that shown by Figure~\ref{fig:angular_distr}. The parameter combinations here are $(B, u_{\rm BLR}, z_0^\gamma, \theta_B) = (50\,{\rm mG}, 50\,\unitU, 0.8\,\unitRBLR, 11\,\degr)$. At this shallower height, where the disk radiation field is far less dense as compared to the height $z_0^\gamma = 0.3\,\unitRBLR$, fewer $\gamma\gamma$ interactions take place. This figure shows a weaker preference of cascade $\gamma$-rays escaping perpendicularly to the plane parallel to the magnetic field, as compared to Figure~\ref{fig:angular_distr}.
\begin{figure}
    \centering
    \includegraphics[width=0.95\linewidth]{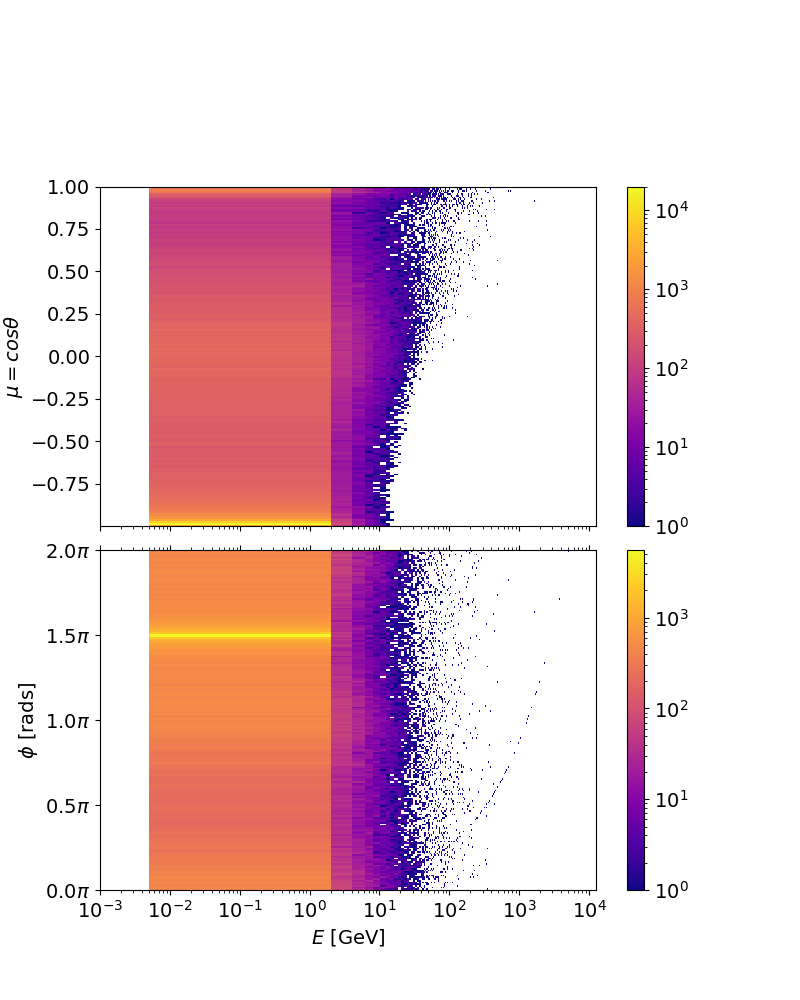}
    \caption{\emph{Upper panel}: 2D histogram plot of angle cosine polar angles versus 
    cascade $\gamma$-ray energy. \emph{Lower panel}: 2D histogram plot of azimuthal angle versus cascade $\gamma$-ray energy.
    }
\label{fig:angular_distrz0.8}
\end{figure}


\bsp	
\label{lastpage}
\end{document}